\documentclass[aps,prd,twocolumn,amssymb,nofootinbib,superscriptaddress,longbibliography]{revtex4-2}

\pdfoutput=1

\usepackage{hyperref}
\usepackage{feynmf}
\usepackage{graphicx}
\usepackage{enumitem}
\usepackage{latexsym}
\usepackage{amsfonts}
\usepackage{amssymb}
\usepackage{mathrsfs}
\usepackage{color}
\usepackage{amsmath}
\usepackage{slashed}
\usepackage{dcolumn}
\usepackage{verbatim}
\usepackage{comment}
\usepackage{float}
\usepackage{multirow}
\usepackage{xspace}
\usepackage{bbm}
\usepackage{bm}
\usepackage{braket}
\usepackage[normalem]{ulem}
\usepackage[dvipsnames]{xcolor}

\newcommand{\GeV}{~\text{GeV}}

\def\triumf{TRIUMF, 4004 Wesbrook Mall, Vancouver, BC V6T 2A3, Canada}
\def\ua{Department of Physics, University of Arizona, Tucson, Arizona 85721, USA}

\begin{document}

\title{High Energy Photon Polarimetry at Lepton Colliders: \\ Quantum Information from Converted Photons}

%\title{Bell Meets Belle}

\author{Carlos Henrique de Lima}
\email{cdelima@triumf.ca}
\affiliation{\triumf}

\author{Navin McGinnis}
\email{nmcginnis@arizona.edu}
\affiliation{\ua}

\author{David McKeen}
\email{mckeen@triumf.ca}
\affiliation{\triumf}

\begin{abstract}
We study high-energy photons produced at a lepton collider that convert into an $e^+e^-$ pair in the detector, as a tool for measuring quantum information observables. We consider single- and double-conversion processes in $e^+e^- \to \gamma\gamma$ and $e^+e^- \to e^+e^-\gamma$. Single conversions enable an \textit{in situ} extraction of the spin-analyzing power, while double conversions probe polarization correlations. Focusing on the Belle-II detector, we show that, depending on the reconstruction resolution of the opening angle of the conversion $e^+e^-$ pair, quantum correlations of the diphoton system can be probed. In particular, measurements of violations of the Bell inequality, quantum discord, concurrence, nonstabilizerness, and steerability with spatially separated GeV-scale photons can be made at high significance.
\end{abstract}

\maketitle

%%%%%%%%%%%%%%%%%%%%%%%%%%%%%%%%%%%%%%%%%%%%%%%
\section{Introduction}
%%%%%%%%%%%%%%%%%%%%%%%%%%%%%%%%%%%%%%%%%%%%%%%

As current collider experiments, such as Belle II and those at the LHC, approach their precision frontier, there is an increased motivation to maximize their physics reach. A large fraction of the information carried by a collision is contained in the spin and polarization of the final state, and in the quantum correlations among them, yet these are rarely measured. Recovering this information from the data already being collected, with the detectors already built, is an economical way to extend the reach of those experiments.

For final states containing photons, this information is encoded in their polarization. Although photons leave no direct track in a detector, their polarization can nevertheless be inferred if they convert into an electron-positron pair in detector material. In the field of a nucleus, the photon converts through the Bethe-Heitler process~\cite{Bethe:1934za} $\gamma +N\rightarrow e^+ e^-+ N$, and the plane of the produced pair is correlated with the photon polarization direction, so that the azimuthal distribution of the pair acts as a polarimeter. This idea has been developed over decades into a mature technique for $\gamma$-ray astronomy, where pair-conversion telescopes measure the polarization of astrophysical sources above the threshold~\cite{Bernard:2013jea,Gros:2016zst,Gros:2016dmp}. At colliders, the same process can be used to obtain additional information on the final state photons, as high-energy photons routinely convert in the silicon of a tracking detector. This approach has been suggested to be used for the measurement of the Higgs CP structure of the coupling to photons~\cite{Bishara:2013vya,Pan:2026usx}, and the study of the photon polarization in $B\to K^\ast\gamma$~\cite{Bishara:2015yta}.

Most approaches at colliders for photon conversion treat them as a background because of how difficult it is to reconstruct the information carried by them~\cite{Jaeckel:2023huy}. The analyzing power of a single conversion is small, of order $A\approx 1/7$; the opening angle of the pair is of order $2m_e/E_\gamma$, and is therefore below a milliradian for$\GeV$ photons. The probability for both photons to convert is of order $10^{-4}$. Measuring the polarization demands both a large sample and an excellent angular resolution on the conversion tracks. Belle~II is well matched to meet these requirements~\cite{Belle-II:2018jsg,Aihara:2024zds}. Its target luminosity of tens of ab$^{-1}$ delivers an enormous number of diphoton events, with each carrying only a few$\GeV$ of energy. The finely segmented vertex detector close to the interaction point allows the reconstruction of small opening angles.

%%%%%%%%%%%%%%%%%%%%%%%%%%%%%%%%%%%%%%%%%%
\begin{figure}[b!]
    \centering
    \includegraphics[width=0.9\linewidth]{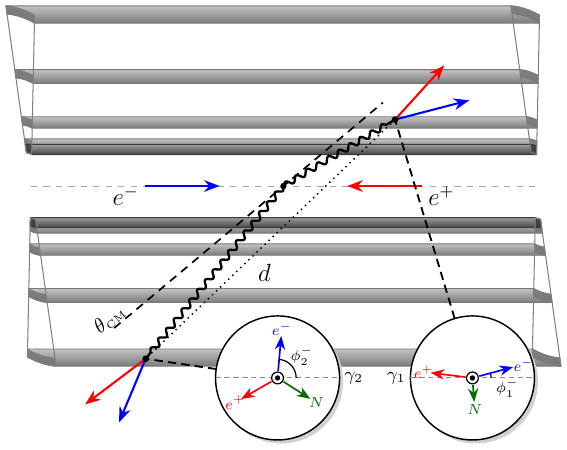}
    \caption{Diphoton conversion into $e^+e^-$ pairs that are spatially separated by a distance $d$ in the Belle II tracker. The insets show projections of the $e^-$ (blue), $e^+$ (red), and nuclear (green) momenta in the planes transverse to the photons' directions (with photons traveling {\em into} the page).}
    \label{fig:schema}
\end{figure}
%%%%%%%%%%%%%%%%%%%%%%%%%%%%%%%%%%%%%%%%%%

At the same time, the study of quantum correlations in collider final states has grown into an active program~\cite{Afik:2020onf,Fabbrichesi:2021npl,Severi:2021cnj,Aoude:2022imd,Fabbrichesi:2022ovb,Afik:2022dgh,Ashby-Pickering:2022umy,Severi:2022qjy,Altakach:2022ywa,Dong:2023xiw,Morales:2023gow,Aoude:2023hxv,Ma:2023yvd,Sakurai:2023nsc,Bernal:2023jba,Han:2023fci,Altomonte:2023mug,Ehataht:2023zzt,Maltoni:2024tul,Aguilar-Saavedra:2024hwd,Blasone:2024dud,Barr:2024djo,Subba:2024mnl,Bernal:2024xhm,CMS:2024pts,Wu:2024asu,Demina:2024dst,Gabrielli:2024kbz,Ruzi:2024cbt,CMS:2024zkc,Du:2024sly,Ravina:2024ard,Cheng:2024rxi,Sullivan:2024wzl,Wu:2024ovc,Ruzi:2024iqu,Altomonte:2024upf,Han:2024ugl,Fabbrichesi:2025ywl,Cheng:2025cuv,Lysak:2025uhk,Han:2025ewp,Fabbrichesi:2025aqp,Aoude:2025ovu,Fabbrichesi:2025psr,Goncalves:2025mvl,Aguilar-Saavedra:2025byk,Aoude:2025jzc,Afik:2025grr,Qi:2025onf,Goncalves:2025xer,Cheng:2025zcf,Bechtle:2025ugc,Abel:2025skj,Wu:2025dds,Aguilar-Saavedra:2025cej,Pei:2025ito,Gu:2025ijz,Yazgan:2025pah,Cheng:2025zaw,Cao:2025xnp,CMS:2025brx,Jolly:2026gpe,Pardos:2026uhw,Guo:2026yhz,Gabrielli:2026tnl,Zhang:2026nwm,Yang:2026uwu,Aguilar-Saavedra:2026wuq,Oussarhan:2026yli,Fang:2026ddi,Goncalves:2026njf,Aoude:2026eeg,Arai:2026jtc,ATLAS:2026nrx,Subba:2026nzs,Wang:2026nls,Zhang:2026wvn,Zhou:2026poo,Goncalves:2026nnx,Liu:2026gxj,Batell:2026bcd,Cheng:2026ktp} with the potential to shed light on or deepen our understanding of enduring puzzles in quantum mechanics~\cite{Einstein:1935rr,Bell:1964kc}. The spin or polarization state of a pair of particles is a bipartite quantum system whose density matrix can be reconstructed from angular distributions, and entanglement and Bell nonlocality have now been studied for top-quark pairs and for a range of other final states~\cite{Afik:2020onf,Barr:2024djo}. The diphoton system is the simplest of these; two polarization qubits and photon pairs have been considered as a target for entanglement and new-physics studies~\cite{Fabbrichesi:2022ovb}. 

Recently, it was proposed to study nearly on-shell diphotons to extract this information at the Belle II experiment~\cite{Cheng:2026ktp}. At leading order, this work studies the same two-qubit system in a complementary way; the measurements we propose utilize real photons, which are often centimeters apart. Furthermore, our treatment of photon conversions can be useful to extract physics from other processes besides diphoton production focused on here. A schematic of the conversion process can be seen in Figure.~\ref{fig:schema}.

In this work, we study in detail the single- and double-photon conversion at Belle II. We quantify the experimental needs for the extraction of polarization information in these processes. Further, we discuss in detail the processes $e^+e^-\to e^+e^- \gamma$ and $e^+e^-\to\gamma\gamma$ where one or both photons convert. We then study the relevant kinematics and perform the reconstruction of quantum information observables. We model the conversion as an imperfect polarimetric measurement and calibrate the analyzing power against a five-dimensional Bethe-Heitler simulation in \textsc{Geant4}~\cite{GEANT4:2002zbu}. We propagate the physics using a simplified description of the vertex detector and the analyzing power, which is sufficient to establish the reach and identify the experimental challenges.

The paper is organized as follows. In Section~\ref{sec:polarimetry}, we introduce the dictionary between quantum information quantities and collider observables and model the photon conversion as an imperfect polarimeter. In Section~\ref{sec:QI}, we define the quantum information observables considered. In Section~\ref{sec:SMandBSM}, we construct the spin density matrix of the photons in $e^+e^-\to\gamma\gamma$ and $e^+e^-\to e^+e^-\gamma$. In Section~\ref{sec:single}, we describe our simplified model of the Belle~II vertex detector and calibrate the analyzing power with the polarized Bethe-Heitler simulation. In Section~\ref{sec:singlepol}, we estimate the reach for the polarization measurement of a single converted photon, which provides a null test of the photon linear polarization in $e^+e^-\to\gamma\gamma$ and an {\em in situ} extraction of the analyzing power from $e^+e^-\to e^+e^-\gamma$ events. In Section~\ref{sec:double}, we estimate the sensitivity of double conversions to the diphoton polarization correlation and to the quantum information observables. We conclude in Section~\ref{sec:conc}.

%%%%%%%%%%%%%%%%%%%%%%%%%%%%%%%%%%%%%%%%%%%%%%%
\section{Photon Polarimetry at Colliders}
\label{sec:polarimetry}
%%%%%%%%%%%%%%%%%%%%%%%%%%%%%%%%%%%%%%%%%%%%%%%

In this section, we describe the correlations between pairs of photons that are produced at a collider, introducing the basis that we use as well as the mapping between field theory quantities and quantum information objects. To gain some insight into the measurement of the quantum correlations of the photon pair, we introduce a toy model of an imperfect photon polarimeter as a simplification of the more complicated conversion process that still captures the essential physics. Our discussion is agnostic about the photon production process, which we will later specify.

%%%%%%%%%%%%%%%%%%%%%%%%%%%%%%%%%%%%%%%%%%%%%%%
\subsection{Quantum Information Dictionary}
\label{sec:dictionary}
%%%%%%%%%%%%%%%%%%%%%%%%%%%%%%%%%%%%%%%%%%%%%%%

A single photon is a two-level system which can be characterized by a $2\times 2$ density matrix $\rho$.
\begin{equation}
\rho=\tfrac12\big(\mathbb{I}+\bm B\cdot\bm\sigma\big)\, ,
\label{eq:rho1}
\end{equation}
where $\sigma^i$ are the Pauli matrices. We choose to work in the circular polarization basis for the photons in which the Stokes vector components $B_x$ and $B_y$ measure the degree of linear polarization along axes $45^\circ$ from one another and $B_z$ measures circular polarization. The polarization state of two photons is a system of two qubits, described by
\begin{equation}
\begin{aligned}
\rho&=\frac14\left(\mathbb{I}_2\otimes\mathbb{I}_2+B^{(1)}_i\,\sigma^i\otimes\mathbb{I}_2\right.
\\
&\quad\quad\quad\quad\left.+B^{(2)}_i\,\mathbb{I}_2\otimes\sigma^i+C_{ij}\,\sigma^i\otimes\sigma^j\right)\, .
\end{aligned}
\label{eq:rho_general}
\end{equation}
Here, $B^{(a)}_i$ is the single-photon Stokes vector of photon $a$ and $C_{ij}$ is the polarization correlation matrix.

The values of the coefficients depend on the production mechanism of the photons, which we specify later. To connect the density matrix to observables we use the $R$-matrix built from the production amplitude $\mathcal{M}_{\lambda_1\lambda_2}$ where $\lambda_{a}$ labels the polarization of photon $a$,
\begin{equation}
R_{\lambda_1\lambda_2\lambda_1^\prime\lambda_2^\prime}=\langle\mathcal{M}_{\lambda_1\lambda_2}\mathcal{M}^\ast_{\lambda_1^\prime\lambda_2^\prime}\rangle\,,\qquad \rho=\frac{R}{{\rm Tr}\, R}\,,
\label{eq:Rmatrix}
\end{equation}
and the angle brackets represent averaging over initial states. The differential cross section is proportional to ${\rm Tr}\, R$.

%%%%%%%%%%%%%%%%%%%%%%%%%%%%%%%%%%%%%%%%%%%%%%%
\subsection{Conversion as an Imperfect Polarimeter}
\label{sec:toy}
%%%%%%%%%%%%%%%%%%%%%%%%%%%%%%%%%%%%%%%%%%%%%%%

The conversion of a single photon to a lepton pair is not a perfect polarization measurement. The orientation of the conversion plane is correlated with the photon polarization, but the correlation is diluted by the nuclear recoil, by the finite opening angle of the pair, and by the multiple scattering and reconstruction of the two tracks. We capture this with a toy model that can be simply mapped onto the polarization description above and which we later calibrate against a full simulation in Sec.~\ref{sec:single}.

To start, we consider a photon traveling in the $\hat{\bm z}$ direction. We orient an imperfect linear polarizer at an angle $\phi$ with respect to the $\bm x$-axis. Using a basis of positively ($\ket{+}$) and negatively ($\ket{-}$) polarized photons, the action of the polarizer can be written
\begin{equation}
P_\phi=\sqrt{\tfrac{1+A}{2}}\,\ket{0_\phi}\bra{0_\phi}+\sqrt{\tfrac{1-A}{2}}\,\ket{1_\phi}\bra{1_\phi}\,,
\label{eq:Pphi}
\end{equation}
with
\begin{equation}
\begin{pmatrix}
\ket{0_\phi} \\
\ket{1_\phi}
\end{pmatrix}=\frac{1}{\sqrt2}
\begin{pmatrix}
e^{-i\phi} & e^{i\phi} \\
-i e^{-i\phi} & ie^{i\phi}
\end{pmatrix}\begin{pmatrix}
\ket{+} \\
\ket{-}
\end{pmatrix}\,.
\end{equation}
This means that the component of a photon parallel to the polarizer is transmitted with intensity $(1+A)/2$ and the orthogonal one with intensity $(1-A)/2$. The associated measurement operator is given by
\begin{equation}
P_\phi^\dagger P_\phi^{\vphantom{\dagger}}=\frac{\mathbb{I}+A\,\hat{\bm n}\cdot\bm\sigma}{2}\,,\qquad \hat{\bm n}=\cos 2\phi\,\hat{\bm x}+\sin 2\phi\,\hat{\bm y}\,.
\end{equation}

The parameter $A$ is the analyzing power. For $A=1$, this is a perfect polarizer along $\phi$; for $A=0$, it is proportional to the identity and carries no polarization information.

The transmitted intensity of a single photon characterized by density matrix $\rho$ incident on the imperfect polarizer as a function of its orientation is
\begin{equation}
\begin{aligned}
\frac{dN}{d\phi}&\propto {\rm Tr}\big[\rho\,P_\phi^\dagger P_\phi^{\vphantom{\dagger}}\big]
\\
&=\frac12\Big[1+A\big(B_x\cos2\phi+B_y\sin2\phi\big)\Big]
\\
&\equiv\frac12\Big[1+AP\cos\big(2\phi+2\phi_0\big)\Big]\, .
\end{aligned}
\label{eq:single_transmission}
\end{equation}
As mentioned above, $B_x$ and $B_y$ describe linearly polarized photons with a phase difference of $45^\circ$ as the polarizer orientation changes. In the last line, as is conventional, we have defined the linear polarization intensity $P = \sqrt{B_x^2+B_y^2}$ and the polarization phase  $\tan2\phi_0 = - B_y/B_x$.

For two photons, each analyzed by its own polarimeter, the joint transmitted intensity is again a trace, now of the two-qubit density matrix against the product of the two measurement effects,
\begin{align} 
&\frac{dN}{d\phi_1 d\phi_2}\propto{\rm Tr}\!\big[\rho\,(P_{\phi_1}^\dagger P_{\phi_1}^{\vphantom{\dagger}})\otimes(P_{\phi_2}^\dagger P_{\phi_2}^{\vphantom{\dagger}})\big] = \\\nonumber
&\frac14\Big[1+A_1\,\bm B^{(1)}\!\cdot\hat{\bm n}_1+A_2\,\bm B^{(2)}\!\cdot\hat{\bm n}_2+A_1A_2\, \hat{\bm{n}}_1^{T}.\bm{C}.\hat{\bm{n}}_2\Big]\,.
\label{eq:joint_full}
\end{align}
The second and third terms on the right-hand side are the single-photon modulations of Eq.~\eqref{eq:single_transmission}, governed by each photon's Stokes vectors $\bm B$; the last term is the genuine correlation, governed by $C_{ij}$, which does not factorize into independent measurements. For an entangled state with unpolarized marginals, $\bm B^{(1,2)}=0$, which we will see below is the case for $e^+e^-\to\gamma\gamma$, the single-photon terms vanish, and the entire signal lives in the joint azimuthal dependence. Using $\hat{\bm n}_i=(\cos2\phi_i,\sin2\phi_i,0)$,
\begin{equation}
\begin{aligned}
\frac{dN}{d\phi_1 d\phi_2} \propto\; 1 + &\frac{A_1 A_2}{2} 
 \Big[(C_{xx}+C_{yy}) \cos 2(\phi_1-\phi_2) \\
&\quad\quad + (C_{xx}-C_{yy}) \cos 2(\phi_1+\phi_2) \\
&\quad\quad + (C_{yx}-C_{xy}) \sin 2(\phi_1-\phi_2) \\
&\quad\quad + (C_{yx}+C_{xy}) \sin 2(\phi_1+\phi_2)
\Big]\, .
\label{eq:joint_all}
\end{aligned}
\end{equation}
Thus, measuring the azimuthal modulation of the two photons allows one to reconstruct the quantum correlations of the system. In Secs.~\ref{sec:single}, \ref{sec:singlepol}, and~\ref{sec:double}, we will discuss how to extract these modulations using converted photons at an $e^+e^-$ collider.

%%%%%%%%%%%%%%%%%%%%%%%%%%%%%%%%%%%%%%%%%%%%%%%
\section{Quantum Information Observables}
\label{sec:QI}

The measurements performed in this work can reconstruct part of the correlation matrix of the diphoton system. From the reconstructed matrix, we can build different quantum information observables. In the following, we discuss several quantum observables that may be extracted from the diphoton polarization state, including entanglement, magic, quantum discord, and quantum steering.

%%%%%%%%%%%%%%%%%%%%%%%%%%%%%%%%%%%%%%%%
\subsection{Entanglement}
\label{sec:ent}
%%%%%%%%%%%%%%%%%%%%%%%%%%%%%%%%%%%%%%%%

Quantum entanglement is the hallmark of quantum correlations. A quantum state is called entangled if it cannot be decomposed into a separable product state. Typically, the amount of entanglement of a bipartite quantum state of qubits is quantified using the concurrence~\cite{Wootters:1997id}
\begin{equation}
    \mathcal{C}[\rho] = \text{max}(0, \lambda_{1}-\lambda_{2}-\lambda_{3}-\lambda_{4}),
\end{equation}
where the $\lambda_{i}$ are the eigenvalues of the operator $\sqrt{\sqrt{\rho}\tilde{\rho}\sqrt{\rho}}$, where $\tilde{\rho} = (\sigma_{2}\otimes\sigma_{2})\rho^{*}(\sigma_{2}\otimes\sigma_{2})$, and $\rho^{*}$ is the entry-wise complex conjugate of $\rho$.

The photon conversion is blind to the circular polarization, and thus the experimentally reconstructed density matrix is incomplete. Consequently, full quantum state tomography is not possible, and quantities such as the concurrence, which require complete knowledge of the density matrix, cannot generally be determined in a model-independent manner. In the specific case of $e^+e^- \rightarrow \gamma \gamma$, the correlation matrix is diagonal, and the circular correlation is constant. We can then reconstruct the full quantum tomography by assuming the predicted value, but these measurements are weaker than the direct reconstruction of the linear correlation.

Even without the full correlation matrix, it is possible to construct a sufficient criterion of entanglement from Bell-type correlation inequalities. In particular, we may define a sufficient condition using the Clauser-Horne-Shimony-Holt (CHSH) inequality~\cite{Clauser:1969ny}. A convenient polarization axis for the construction of the CHSH condition is the Cartesian basis, such that the inequality becomes~\cite{Severi:2021cnj}
\begin{align}
    \mathcal{E}_{\pm}= | C_{xx} \pm C_{yy} | < \sqrt{2}
        \label{eq:CHSH_simp}
\end{align}

Violating this inequality would imply $\mathcal{C}[\rho]>0$. It should be noted that satisfying Eq.~\ref{eq:CHSH_simp} with a specific choice of polarization projection directions does \textit{not} imply a test of local realism. Rather, violating the CHSH inequality shows that given the reconstructed values of $C_{ij}$, entanglement of the quantum state $\rho$ can be established for the chosen polarization directions.

\subsection{Magic}
\label{sec:magic}

Magic quantifies the extent to which a quantum state cannot be efficiently simulated by stabilizer techniques and is therefore regarded as a computational resource for fault-tolerant quantum computation. Recently, magic has also emerged as a useful measure of non-classicality in relativistic scattering processes and collider observables~\cite{White:2024nuc,Liu:2025qfl,Aoude:2025jzc,Busoni:2025dns,Gargalionis:2025iqs,Liu:2025bgw,Cheng:2025zaw,Afik:2026pxv,Gargalionis:2026onv}. Stabilizer states form a subset of quantum states that admit an efficient classical description~\cite{Gottesman:1998hu}. 

A pure $n$-qubit stabilizer state is defined as the unique simultaneous $+1$ eigenstate of a maximal set of $n$ independent, commuting Pauli operators. Equivalently, every stabilizer state may be prepared from a computational basis state using only Clifford operations defined by
\begin{equation}
    \mathcal{C}_{n} = \{U\in U(2^{n}): UPU^{\dagger} = e^{i\theta}P^{\prime}\},
\end{equation}
where $P,P^{\prime}$, are members of the Pauli group
\begin{equation}
\mathcal{P}_n=
\left\{
P_1\otimes\cdots\otimes P_n
\,\middle|\,
P_i\in\{I,\sigma_x,\sigma_y,\sigma_z\}
\right\},
\end{equation}
and $\theta\in\{0,\pi/2,\pi,3\pi/2\}$. Pure states that are not stabilizer states, or, more generally, mixed states that lie outside the convex hull of stabilizer states, are said to possess quantum magic.

Measures of quantum magic quantify the extent to which a quantum state fails to admit a stabilizer description.  We quantify the non-stabilizerness of the diphoton polarization state using the Rényi magic monotone,
\begin{equation}
    \mathcal {M}_{q}= \frac{1}{1-q}\log_{2}\left(\frac{\sum_{P\in\mathcal{P}_{n}}\braket{\psi|P|\psi}^{2q}}{\sum_{P\in\mathcal{P}_{n}}\braket{\psi|P|\psi}^{2}}\right)
\end{equation}
For pure states, the Rényi magic may be computed directly from the expectation values of the Pauli group. Since the diphoton polarization state reconstructed from photon conversion is generally mixed, we employ the corresponding mixed-state extension evaluated from the reconstructed density matrix.

\subsection{Quantum Discord}
\label{sec:discord}
Quantum discord quantifies correlations that are fundamentally quantum in origin. A bipartite state may possess nonzero discord even when it is separable, reflecting the fact that local measurements can irreversibly disturb the state despite the absence of entanglement. Quantum discord therefore provides a more general characterization of quantum correlations than entanglement.

Once the Fano coefficients have been extracted from the Stokes parameters, discord may be computed directly from the reconstructed density matrix by performing a numerical optimization over local projective measurements. In particular, the quantum discord of the $(1)$ photon is given by
\begin{equation}
    D_{(1)}(\rho)=S(\rho_{(2)})-S(\rho)+\min_{\hat{\bm{n}}}(p_{+\hat{\bm{n}}}S(\rho_{+\hat{\bm{n}}})+p_{-\hat{\bm{n}}}S(\rho_{-\hat{\bm{n}}})),
    \label{eq:discord_def}
\end{equation}
where $S(\rho)=-\text{Tr}[\rho\log_{2}(\rho)]$ is the Von Neumann entropy, and

\begin{flalign}
    p_{\pm\mathbf{\hat{n}}}=&\frac{1\pm \mathbf{\hat{n}}\cdot\bm{B}^{(2)}}{2},\\
\rho_{\pm\mathbf{\hat{n}}}=&\frac{\mathbb{I}_2+\bm{B}^{(1)}_{\pm\mathbf{\hat{n}}}\cdot\bm{\sigma}}{2},\\
    \bm{B}^{(1)}_{\pm \mathbf{\hat{n}}}=&\frac{\bm{B}^{(1)}\pm \bm{C} .\mathbf{\hat{n}}}{1\pm\mathbf{\hat{n}}\cdot\bm{B}^{(2)}}.
\end{flalign}
Quantum discord of the $(2)$ photon is similarly obtained via the replacements $(1)\leftrightarrow (2)$.

\subsection{Steering}
\label{sec:steering}
Quantum steering characterizes the ability of one subsystem to remotely influence the set of quantum states accessible to a second subsystem through local measurements. Every Bell-nonlocal state is steerable, and every steerable state is entangled. 

For two-qubit systems, steerability can be tested using a variety of \textit{steering inequalities}\cite{Skrzypczyk:2014stn,Jevtic:2014icx} or equivalent geometric criteria based on the density matrix. In particular, for an unpolarized state $\bm B^{(1,2)}=0$, a necessary and sufficient condition for a quantum state to be steerable is given by the inequality
\begin{equation}
   \mathcal{S} = \frac{1}{2\pi}\int d\hat{\bm{n}} \sqrt{\hat{\bm{n}}^{T}\cdot \bm{C}^{T}.\bm{C}.\hat{\bm{n}}} > 1
\end{equation}
Throughout this work, we quantify steering using this criterion. 

%%%%%%%%%%%%%%%%%%%%%%%%%%%%%%%%%%%%%%%%%%%%%%%%%%%%%%%%%%%%%%%%
\begin{figure*}[t!]
    \centering
    \includegraphics[width=0.4\linewidth]{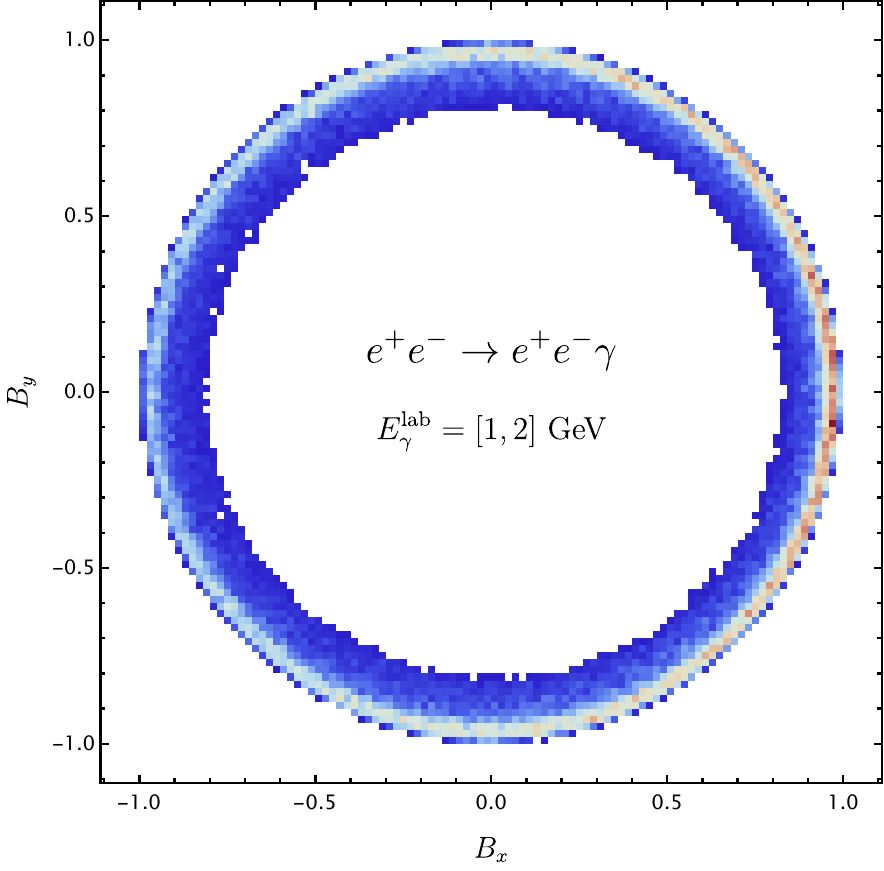} \, \,
    \includegraphics[width=0.4\linewidth]{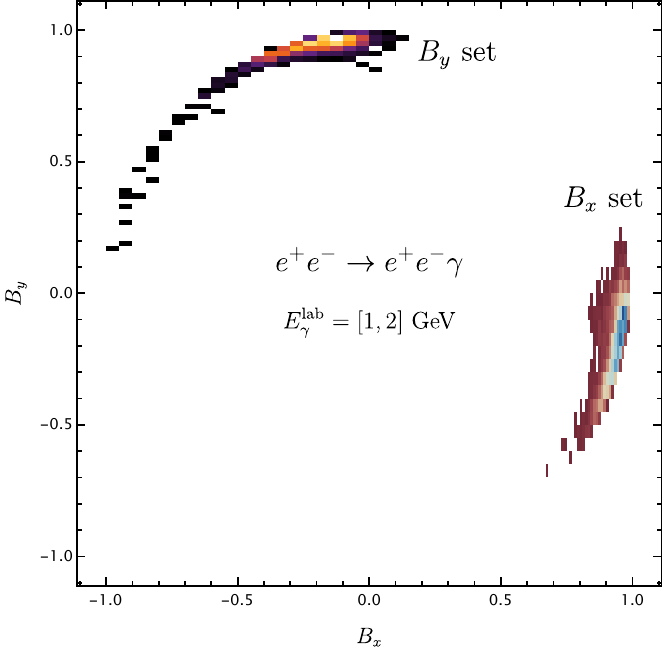}
    \caption{Normalized 2D Histograms of the linear Stokes parameters of the photon from the $e^+e^-\gamma$ process before (\textbf{left}) and after (\textbf{right}) event selection. The selection creates samples with high purity in $B_x$ or $B_y$.}
    \label{fig:linear}
\end{figure*}
%%%%%%%%%%%%%%%%%%%%%%%%%%%%%%%%%%%%%%%%%%%%%%%%%%%%%%%%%%%%%%%%%

%%%%%%%%%%%%%%%%%%%%%%%%%%%%%%%%%%%%%%%%%%%%%%%
\section{Spin Density Matrix for $\gamma\gamma$ and $e^+ e^- \gamma$}
\label{sec:SMandBSM}
%%%%%%%%%%%%%%%%%%%%%%%%%%%%%%%%%%%%%%%%%%%%%%%

In this section, we construct the spin density matrix for the photons in the processes explored in this work, namely $e^+e^-\to\gamma\gamma$ and $e^+e^-\to e^+e^-\gamma$. Since the former is a two-body process while the latter is a three-body process, they exhibit significantly different kinematics. In each case, we discuss the kinematic selections that maximize the linear polarization of the photons and the correlations between the photons for the $\gamma\gamma$ final state.

%%%%%%%%%%%%%%%%%%%%%%%%%%%%%%%%%%%%%%%%%%%%%%%
\subsection{$\gamma\gamma$}
\label{sec:gammagamma}
%%%%%%%%%%%%%%%%%%%%%%%%%%%%%%%%%%%%%%%%%%%%%%%

Diphoton production is comparatively simple because of its two-body kinematics. In the center of mass frame, the photons are monochromatic, with each having energy $\sqrt s/2$, which at a $B$-factory like Belle II is about $5.3~{\rm GeV}$. The density matrix for $e^+e^-\to\gamma\gamma$  is particularly simple, and we obtain the following analytic expressions for the Stokes vectors and correlations,
\begin{align}
    \bm B^{(1)} &= \bm B^{(2)} = \left(\frac{8m_e^2}{s}\frac{1}{1+\cos^2\theta^\ast},0,0\right)\, ,\label{eq:Bgammagamma} \\
    C_{xx} & = C_{yy} = -\frac{\sin^2\theta^\ast}{1+\cos^2\theta^\ast}+{\cal O}\left(\frac{m_e^2}{s}\right)\, , \\
    C_{zz} & = -1+{\cal O}\left(\frac{m_e^2}{s}\right)\, , \\
    C_{ij} &= 0~{\rm for}~i\neq j\, .
\end{align}
Note that the linear polarization is helicity-suppressed, which, at Belle II energies, is beyond the experimental reach. Another important observation is that, because of the simplicity of the correlation matrix, the concurrence of this system is directly $\mathcal{C}[\rho]=|C_{xx}|=|C_{yy}|$. 

The Belle II experiment analyzes $e^+e^-$ collisions with asymmetric beam energies so that the lab frame is boosted with respect to the center of mass. The photons from $e^+e^-\to\gamma\gamma$ are emitted back-to-back in the center of mass with polar angles
\begin{equation}
\cos\theta^{\ast 1,2}=\pm\cos\theta^\ast \,.
\end{equation}
so that the lab-frame polar angles of the two photons are
\begin{equation}
\cos\theta_{\rm lab}^{1,2}=\frac{\beta_{\rm cm}\pm\cos\theta^*}{1\pm\beta_{\rm cm}\cos\theta^*}\,,
\label{eq:boost}
\end{equation}
with $\beta_{\rm cm}=0.27$, the boost of the lab frame, so the two photons carry energies between $4.1$ and $6.9\GeV$ across the acceptance (with a one-to-one correspondence between the energy and $\theta_{\rm lab}$). The differential cross-section is
\begin{equation}
\frac{d\sigma}{d\cos\theta^*}=\frac{{\rm Tr}\, R}{32\pi s}=\frac{2\pi\alpha^2}{s}\frac{1+\cos^2\theta^*}{\sin^2\theta^*}\, .
\label{eq:dsigma}
\end{equation}
Using $\sqrt s=10.58~{\rm GeV}$, $\alpha(\sqrt s)\simeq1/132$, and requiring that both photons satisfy $17^\circ<\theta_{\rm lab}<150^\circ$ gives a fiducial cross section of $\sigma_{\rm fid}=2.9~{\rm nb}$. In other words, with $100~{\rm fb}^{-1}$ of data, there are $2.9\times10^8$ diphoton events in the acceptance. 

Note that as a result of parity invariance, the cross-section and the linear correlation are inversely related,
\begin{equation}
\frac{d\sigma}{d\cos\theta^*}\propto \frac{1}{\left|C_{xx,yy}(\theta^*)\right|}\,.
\end{equation}
This means that while central photons carry most of the correlation information, they are also the least likely configuration to be emitted.

%%%%%%%%%%%%%%%%%%%%%%%%%%%%%%%%%%%%%%%%%%%%%%%
\begin{figure*}[t!]
    \centering
    \includegraphics[width=0.475\linewidth]{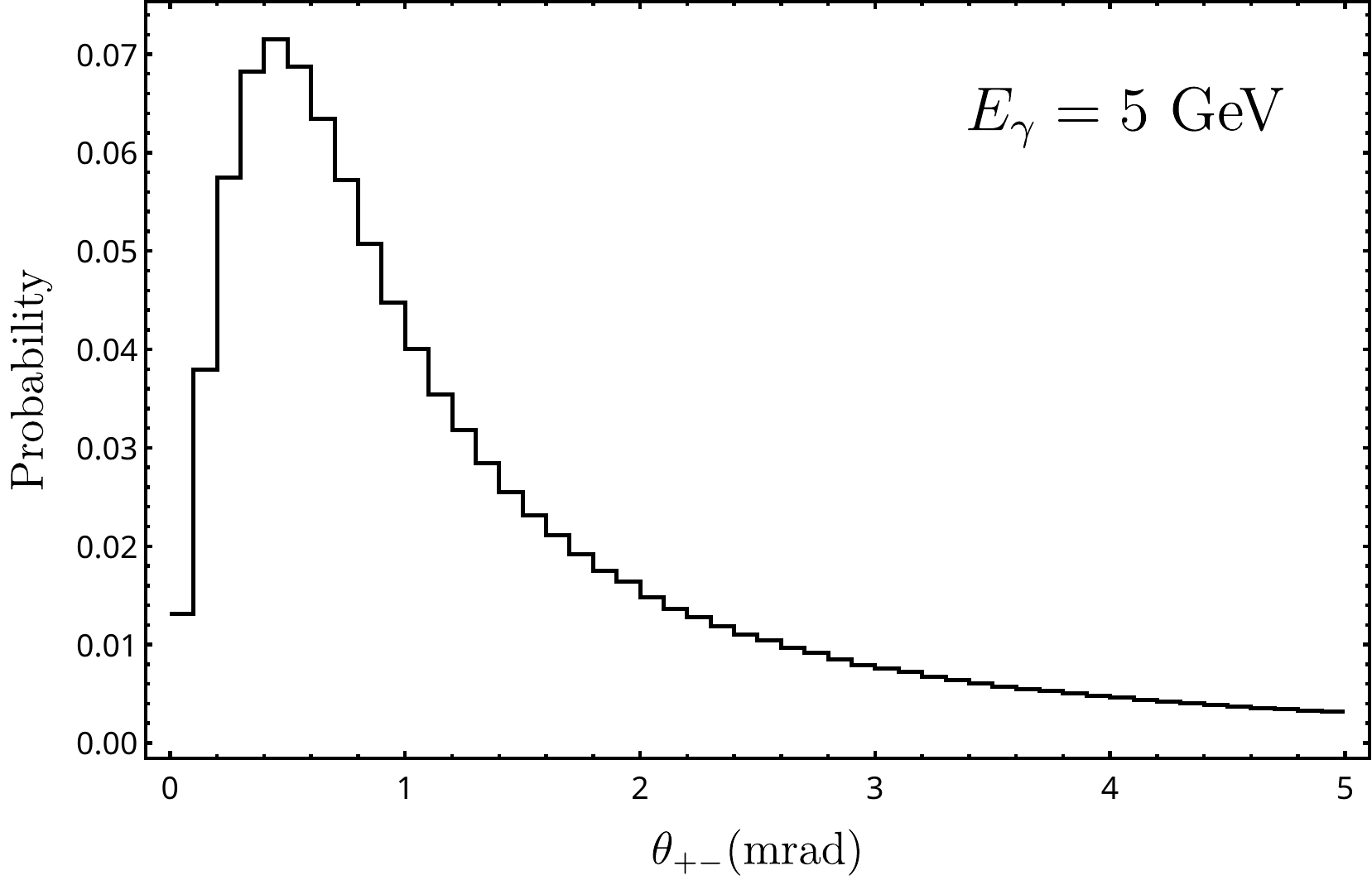} \, \,
    \includegraphics[width=0.475\linewidth]{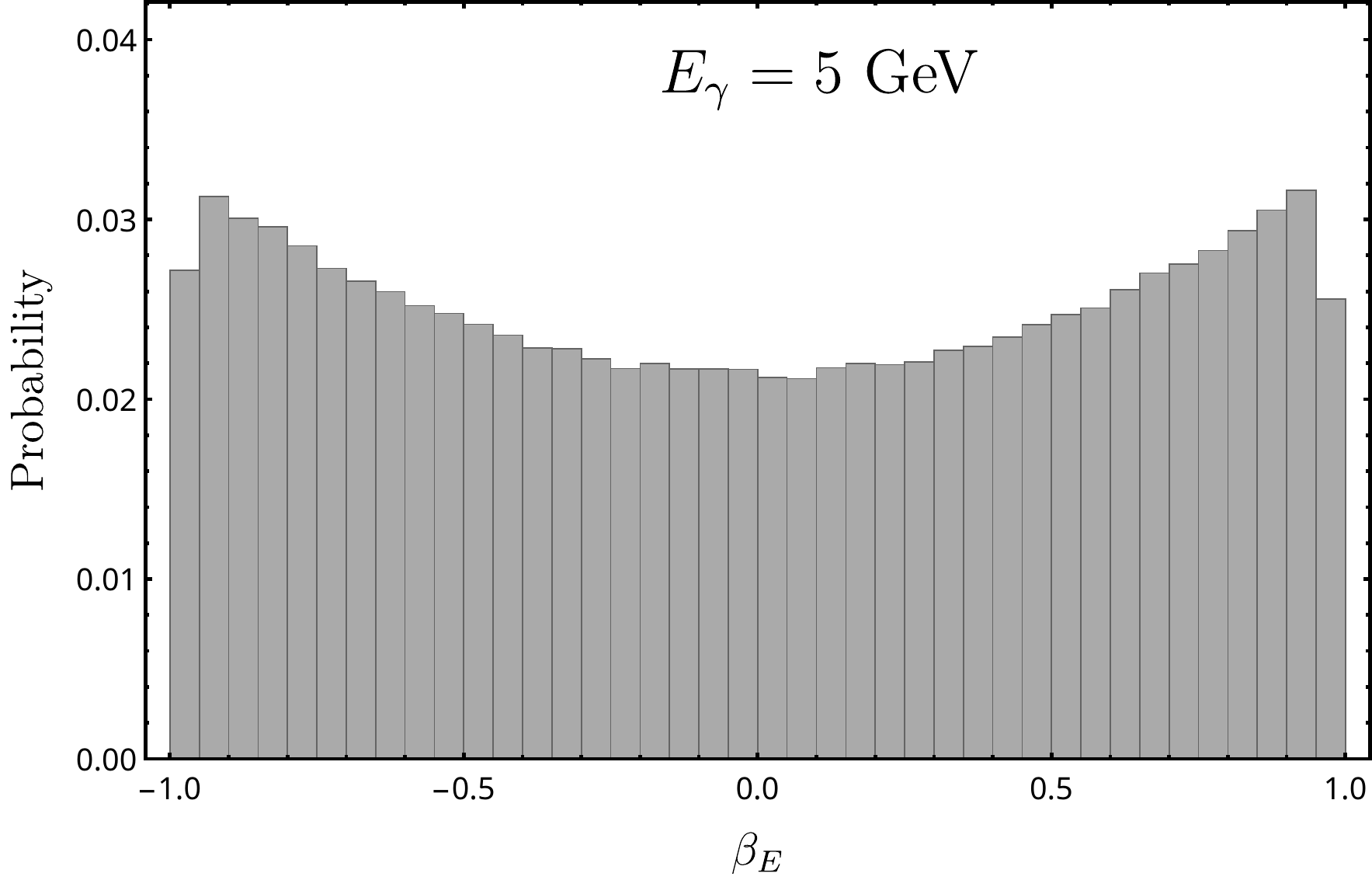}
    \caption{Kinematics of a $5\GeV$ photon conversion from the Bethe-Heitler simulation. Left: the distribution of the $e^+e^-$ opening angle $\theta_{\pm}$, peaked near $2m_e/E_\gamma\sim0.2$~mrad. Right: the distribution of the energy balance $\beta_E=(E_+-E_-)/(E_++E_-)$, which is broad with a mild preference for asymmetric sharing.}
    \label{fig:kinematics}
\end{figure*}
%%%%%%%%%%%%%%%%%%%%%%%%%%%%%%%%%%%%%%%%%%%%%%%

%%%%%%%%%%%%%%%%%%%%%%%%%%%%%%%%%%%%%%%%%%%%%%%
\subsection{$e^+ e^- \gamma$}
\label{sec:eegamma}
%%%%%%%%%%%%%%%%%%%%%%%%%%%%%%%%%%%%%%%%%%%%%%%

The $e^+ e^- \gamma$ final state has much more complicated kinematics than diphoton production. Because it is three-body, the spin density matrix of the photon depends on the kinematic information of the prompt $e^+e^-$ pair. Although this process cannot offer insight into the entanglement of photons at macroscopic distances, it can be used to obtain an {\em in situ} measurement of the analyzing power of the Belle II detector.

We numerically generate $e^+ e^- \to e^+ e^- \gamma$ events with asymmetric beams at $\sqrt{s}=10.58~{\rm GeV}$ corresponding to the Belle II configuration. At each phase space point, we evaluate the helicity amplitudes, mapping onto the single photon density matrix of Eq.~(\ref{eq:rho1}). The important quantities are $B_x$ and $B_y$, which correspond to the linear polarizations that induce different modulations when the photon later converts. Our analysis considers photons with lab frame energies $1~{\rm GeV}<E_\gamma<2~{\rm GeV}$ and all final states to be within the tracker acceptance of $17^\circ<\theta<150^\circ$. This results in a fiducial cross section of $280~{\rm pb}$. 

The process produces linearly polarized photons but with different phases. The degree of linear polarization increases with decreasing photon energy, which is why we restrict to photon energies between $1$ and $2~{\rm GeV}$.~\footnote{We do not include lower energies than $1~{\rm GeV}$ to avoid issues with conversion $e^\pm$ rescattering, which can affect the reconstruction of their kinematics.} In the left panel of Fig.~\ref{fig:linear} we show the resulting $B_x$ and $B_y$ values of our sample of events. The large degree of linear polarization is apparent since the distribution is well separated from the origin.~\footnote{A complementary sample can be constructed with the invariant mass of the prompt $e^+e^-$ pair restricted to be small. This is dominantly populated by nearly on-shell virtual photons. In this limit, the process is very closely related to $\gamma\gamma$ production, with the real photon's center-of-mass energy approaching $\sqrt s/2$, which does not feature linear polarization as seen in Sec.~\ref{sec:gammagamma} and in Ref.~\cite{Cheng:2026ktp}.}

The extraction of the polarization measurement performed in this work relies on a good selection of these events. We would like to create two sets of samples that are of high purity in $B_x$ and in $B_y$ separately. Because of the complicated kinematics, this information is correlated between all three particles. As our goal is to highlight the extraction, not necessarily the optimization of the kinematic selection, we select those events using a simple neural network. This trades interpretability for the possibility of applying cuts on a highly nonlinear selection variable. The neural network~\cite{Janiesch_2021} uses as input the four vectors with labels $B_x$ and $B_y$ of that kinematic point. We then propagate this information through three hidden layers, each followed by Ramp (ReLU) activations~\cite{2018arXiv180308375A}, enabling nonlinear feature extraction. A final linear layer maps to two outputs, followed by a tanh activation that constrains predictions to the interval (-1, 1). The training uses a mean squared error loss function and outputs two scores $S_x$ and $S_y$.

The score of the regression model provides good separation of the data. We perform two selections that create high-quality events. The $B_x$ selection chooses scores $S_x>0.9$ and $S_y<0$ and the selection for $B_y$ selects scores  $S_x<0$ and $S_y>0.9$. These selections create well-separated samples with means $(B_x,B_y) = (0.91,-0.23)$ and $(B_x,B_y) = (-0.21,0.91)$, with cross sections of $\sigma_{1} = 13~\text{pb}$ and $\sigma_{2} = 10~\text{pb}$, respectively. The $B_x$ and $B_y$ values of these selections can be seen in the right panel of Fig.~\ref{fig:linear}. This is by no means optimized, but it is enough for a high-precision measurement of the analyzing power using single conversion, as we will show later.

%%%%%%%%%%%%%%%%%%%%%%%%%%%%%%%%%%%%%%%%%%
\begin{figure*}[t!]
    \centering
    \includegraphics[width=0.48\linewidth]{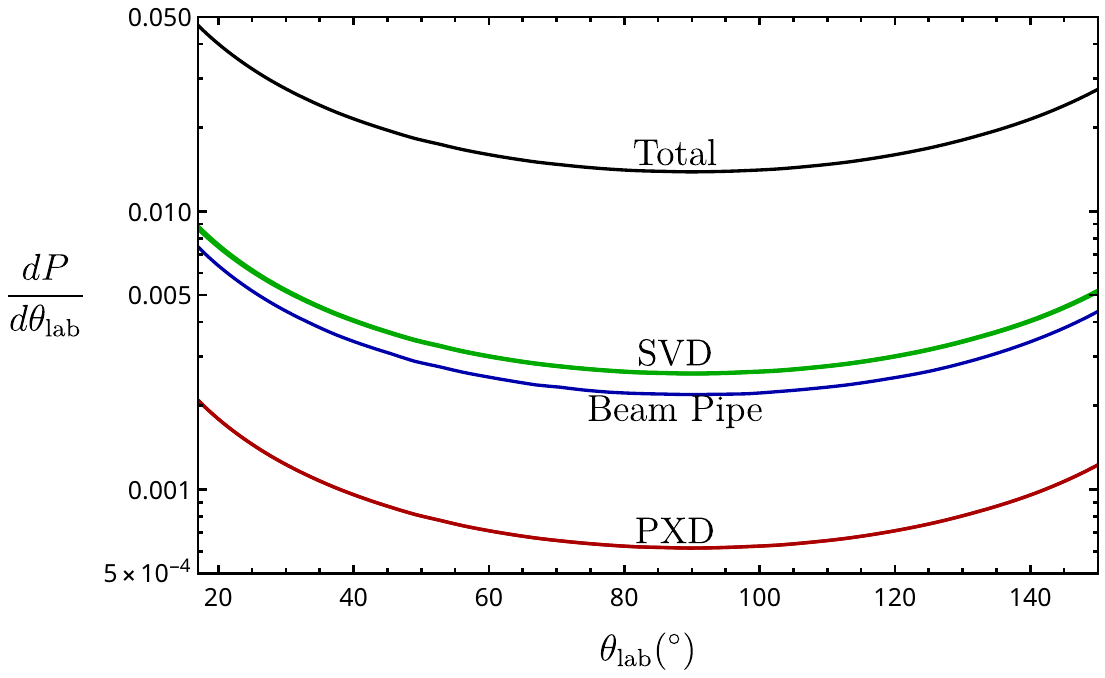} \, \,
    \includegraphics[width=0.47\linewidth]{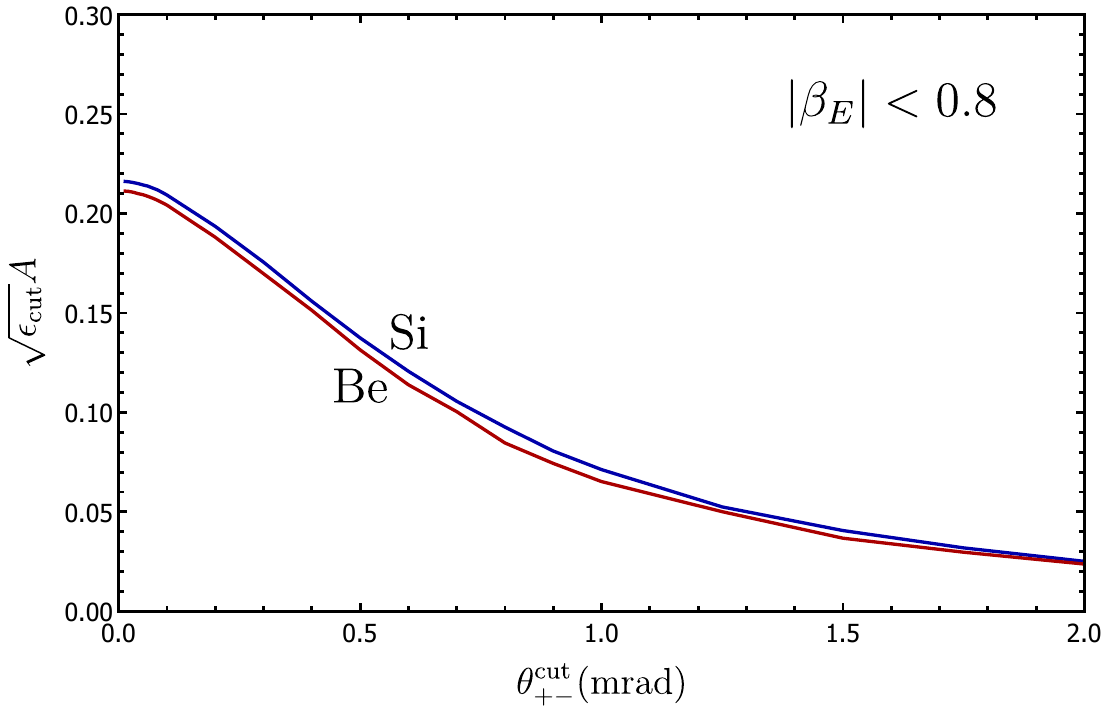}
    \caption{ \textbf{Left:} Single-photon conversion probability as a function of the lab polar angle $\theta_{\rm lab}$, broken down by detector element. \textbf{Right:} Figure of merit $\sqrt{\epsilon_{\rm cut}}A$ as a function of the opening-angle selection $\theta_{\pm}^{\rm cut}$, for conversions in silicon and beryllium at fixed energy $E_\gamma = 5 $ GeV and energy balance $|\beta_E|<0.8$.}
    \label{fig:convprobandcut}
\end{figure*}
%%%%%%%%%%%%%%%%%%%%%%%%%%%%%%%%%%%%%%%%%%

%%%%%%%%%%%%%%%%%%%%%%%%%%%%%%%%%%%%%%%%%%%%%%%
\section{Photon Conversion at Belle II}
\label{sec:single}
%%%%%%%%%%%%%%%%%%%%%%%%%%%%%%%%%%%%%%%%%%%%%%%

The photons created in the $e^+e^-$ collisions propagate through the detector, and a fraction is converted in the beryllium beam pipe and the silicon layers of the vertex detector. The plane formed by the conversion pair retains information about the linear polarization of the parent photon, so the azimuthal orientation of the pair, reconstructed from the two tracks, is the observable that carries information about the density matrix and the entanglement it encodes.

In this section, we characterize single conversions and fix the ingredients needed to describe double conversions in Sec.~\ref{sec:double}. We describe a simplified model of the vertex detector and the $\gamma N\to e^+e^- N$ Bethe-Heitler process, and we calibrate the analyzing power against a five-dimensional polarized event generator in \textsc{Geant4} that reproduces the nuclear recoil and the full differential cross section. We quantify the tradeoff between the analyzing power and the retained efficiency as a function of the track resolution, and we identify the high-quality conversions, which occur early enough for the pair to separate across the tracker and enable the cleanest measurement. 

%%%%%%%%%%%%%%%%%%%%%%%%%%%%%%%%%%%%%%%%%%%%%%%
\subsection{Detector Modeling and Simulation of Pair Conversion}
\label{sec:setup}
%%%%%%%%%%%%%%%%%%%%%%%%%%%%%%%%%%%%%%%%%%%%%%%

We adopt a simplified description of the Belle~II vertex detector that retains the features relevant for a phenomenological study while avoiding a full detector simulation. We model the vertex detector based on its dominant material budget following~\cite{Jaeckel:2023huy,Belle-II:2018jsg} as seven concentric cylindrical shells: the beryllium beam pipe and the six silicon layers of the pixel and strip detectors, with the radii, thicknesses, and materials listed in Table~\ref{tab:layer_conversions}. A photon emitted at polar angle $\theta_{\rm{lab}}$ crosses each shell with a path length proportional to $1/\sin\theta_{\rm{lab}}$, and converts in each with the probability given in the table. We neglect the inner wall of the drift chamber and the support structures, which contribute little to the conversion probability and lie outside the region where a precise vertex can be reconstructed.

%%%%%%%%%%%%%%%%%%%%%%%%%%%%%%%%%%%%%%%%%%%%%%%%%%%%
\begin{table}[h!]
\centering
\begin{tabular}{l | c | c | c | c}
\hline\hline
Layer & Material & $r$ [mm] & $z$ [mm] & Conversions [\%] \\
\hline
Beam Pipe L$_0$ & Be & $10.0$  & $1.000$ & $16.36$ \\
PXD L$_1$   & Si & $14.0$  & $0.075$ & $4.43$  \\
PXD L$_2$   & Si & $22.0$  & $0.075$ & $4.43$  \\
SVD L$_3$   & Si & $38.0$  & $0.320$ & $18.83$ \\
SVD L$_4$   & Si & $80.0$  & $0.320$ & $18.74$ \\
SVD L$_5$   & Si & $104.0$ & $0.320$ & $18.65$ \\
SVD L$_6$   & Si & $135.0$ & $0.320$ & $18.56$ \\
\hline\hline
\end{tabular}
\caption{The simplified Belle~II vertex detector: material, radius $r$, thickness $z$, and the geometric the fraction of conversions occurring in each layer.}
\label{tab:layer_conversions}
\end{table}
%%%%%%%%%%%%%%%%%%%%%%%%%%%%%%%%%%%%%%%%%%%%%%%%%%%%

We do not simulate the Belle~II trigger, reconstruction, or detector-level backgrounds. Our goal is to establish the reach and to identify the experimental challenges, the most important of which is the track angular resolution that controls the analyzing power. More importantly, the lower limit of the angular resolution, where there is a critical loss on the reconstruction, is the feature that most affects the experimental reach. Detector smearing of the reconstructed azimuthal angle by a few degrees does not degrade the analysis reach significantly.

A photon converts in the field of a nucleus through the Bethe-Heitler process $\gamma N \to e^+e^- N$. The differential cross section depends on five kinematical variables of the pair: the energy of the electron (or positron), the two polar production angles, and the two azimuthal angles. The azimuthal distributions contain information about the initial photon polarization~\cite{Bernard:2013jea,Bernard:2018hwf}. For a linearly polarized photon, the cross section contains a term proportional to $P\cos2(\varphi+\varphi_0)$, with $\varphi$ the azimuth of the pair and $\varphi_0$ the photon polarization angle, and this modulation is what the tracker measures. The nuclear recoil dilutes the spin-analyzing power, and the bisection of the two azimuths, $\phi_{\pm}= (\phi_++\phi_-)/2$, is the most robust variable against this effect~\cite{Gros:2016dmp}, and we use it for our analysis. A schematic of the nuclear recoil effect on the azimuthals can be seen in the insets of Fig.~\ref{fig:schema}.

%%%%%%%%%%%%%%%%%%%%%%%%%%%%%%%%%%%%%%%%%%%%%%%%%%%%
 \begin{figure*}[t!]
    \centering
    \includegraphics[width=0.475\linewidth]{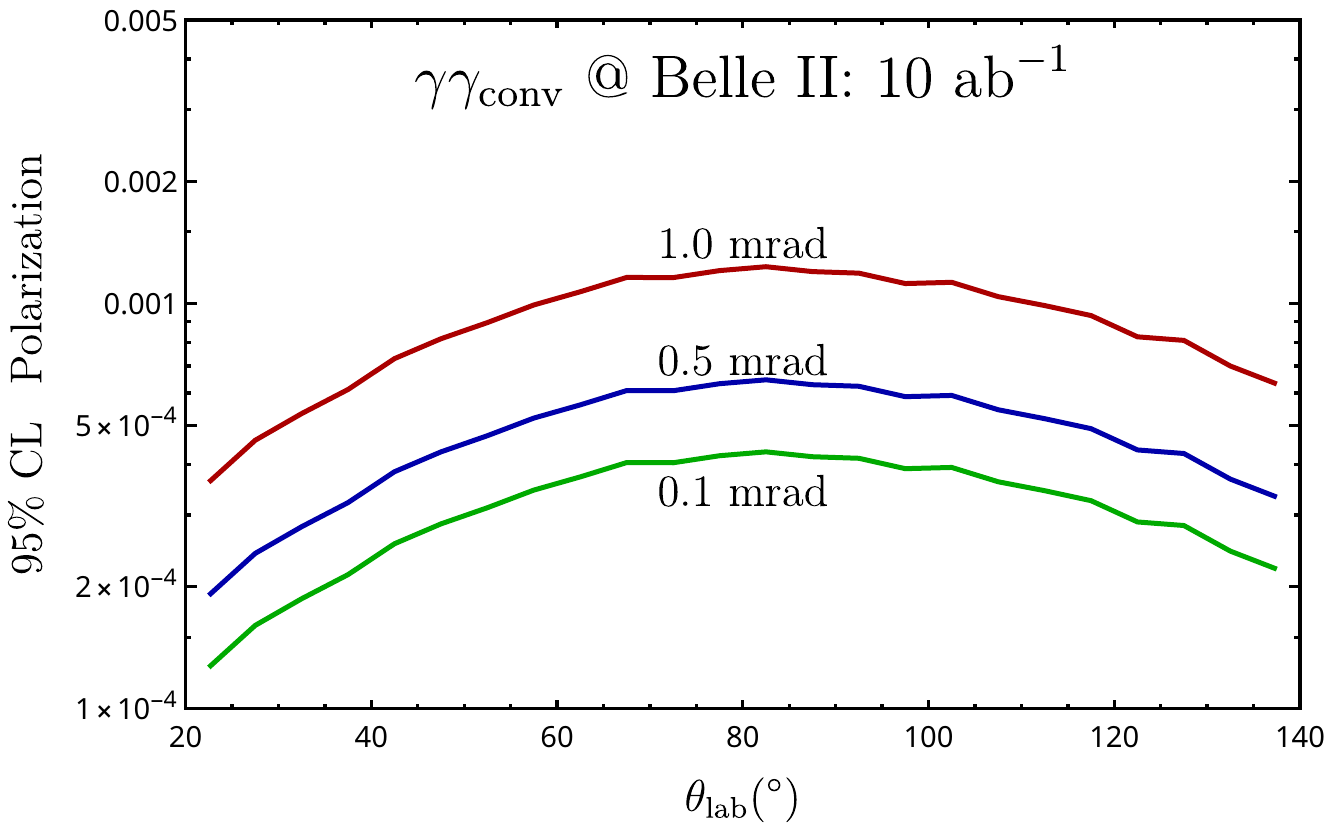} \, \,
    \includegraphics[width=0.45\linewidth]{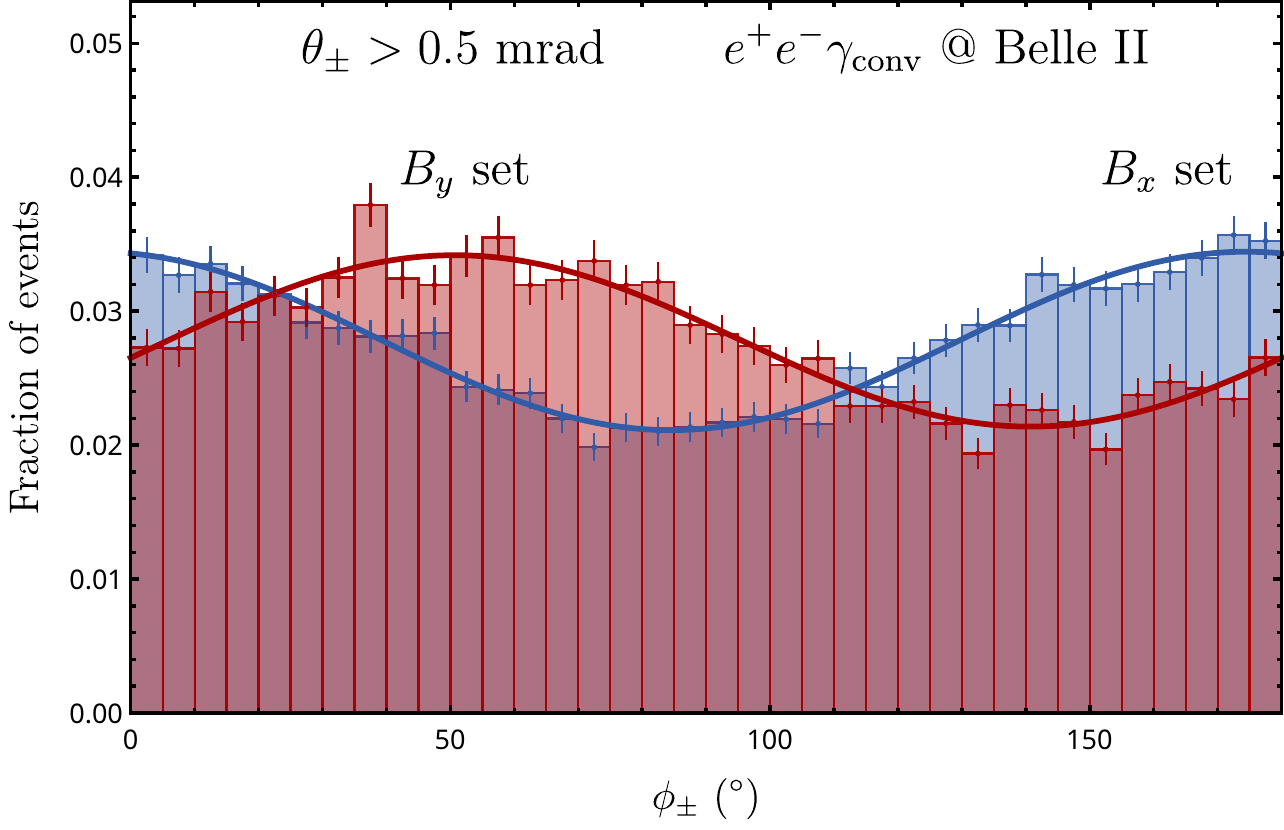}
    \caption{\textbf{Left:} Single-conversion null test: the $2\sigma$ sensitivity to the photon linear polarization $P$ assuming $\phi_0=0$ ($B_x$ only) as a function of the lab angle at $\mathcal{L}=10$ ab$^{-1}$, for the three track resolutions of Table~\ref{tab:cuts}. \textbf{Right:} Normalized azimuthal distribution of the pair bisection angle $\phi_{\pm}$ for the two selections ($B_x$ or $B_y$ enriched) in (blue, red), assuming the minimal angular resolution for the opening angle of 0.5~mrad. Histograms show the expected per-bin fraction of the single conversion with error bars considering 100 fb$^{-1}$ of data.}
    \label{fig:single_conv}
\end{figure*}
%%%%%%%%%%%%%%%%%%%%%%%%%%%%%%%%%%%%%%%%%%%%%%%%%%%%

The azimuthal modulation is largest for symmetric energy sharing between $e^-$ and $e^+$ and for pairs with a small opening angle. The opening angle of the pair is of order $2m_e/E_\gamma$, which for a $5\GeV$ photon is about $0.2$ mrad, so the two tracks are nearly collinear, and the azimuthal orientation of the pair must be reconstructed from tracks separated by a small angle. The achievable analyzing power is therefore extremely sensitive to the track angular resolution. These kinematic features are shown in Fig.~\ref{fig:kinematics}.

We obtain the analyzing power in two steps. We generate the conversion with a five-dimensional polarized Bethe-Heitler event generator in \textsc{Geant4}~\cite{Bernard:2018hwf} that samples the full differential cross section including the nuclear recoil. For the simulation, the only nontrivial dependence comes from the detector material and photon energy. In the setup that we explore, the photons are high-energy, and the analyzing power roughly becomes independent of energy.

A high-energy photon traversing a thickness $X$ of material of radiation length $X_0$ converts with probability
\begin{equation}
p_{\gamma N\to e^+e^- N}(\sin\theta_{\rm{lab}})=1-\exp\!\left(-\frac{7}{9}\frac{X}{X_0\sin\theta_{\rm{lab}}}\right)\,,
\label{eq:pconv}
\end{equation}
where the factor $7/9$ is the asymptotic pair-production fraction of the radiation length and the $1/\sin\theta_{\rm{lab}}$ accounts for the path length through a cylindrical layer at polar angle $\theta_{\rm{lab}}$. For silicon, $X_0=93.7$~mm, so each millimeter of silicon converts a $5\GeV$ photon with probability $8.3\times10^{-3}$. The per-layer conversion fractions are given in Table~\ref{tab:layer_conversions}, the total single-photon conversion probability, integrated over the material and averaged over the fiducial acceptance, is roughly $2.2\times10^{-2}$, and its dependence on the lab angle is shown in the left panel of Fig.~\ref{fig:convprobandcut}. Note that because the probability of conversion is not flat, it distorts distributions as a function of their $\theta_{\rm{lab}}$.

The azimuthal angle of the conversion plane is reconstructed from the two track directions, whose transverse separation grows with the distance from the conversion point to the outer hits. A pair created in the third layer with an opening angle of $0.5$~mrad reaches the outermost layer with a separation of roughly $50~\mu$m, while $0.1$~mrad gives about $10~\mu$m. These conversions occurring at $L_3$ or earlier have a higher quality, where the lever arm to the outer layers is long enough to split the pair into distinct hits. From the per-layer fractions of Table~\ref{tab:layer_conversions}, we expect such conversions to be $44\%$ of all conversions.

%%%%%%%%%%%%%%%%%%%%%%%%%%%%%%%%%%%%%%%%%%%%%%%
\subsection{Polarization Measurement and Analyzing Power}
\label{sec:Acuts}
%%%%%%%%%%%%%%%%%%%%%%%%%%%%%%%%%%%%%%%%%%%%%%%

The analyzing power realized in practice depends on the selection applied to the reconstructed pair. A symmetric, small-opening pair carries more polarization information than an asymmetric, wide-open one, but requiring such pairs lowers the efficiency. To quantify the tradeoff of different kinematic cuts, we can estimate the reach required to be able to differentiate the azimuthal modulation. 

Consider the number of converted photons to be $N_{\gamma\text{-}\rm{conv}}$ without any cut on the converted pair. Then, consider that a given cut on the converted pair reduces the number of events by $\epsilon_{\rm{cut}}$ and with that selection, the analyzing power is given by $A(\epsilon_{\rm{cut}})$. Since not all $e^+e^-$ pairs are oriented with respect to the photon polarization plane, we have to perform a measurement over an ensemble of conversion pairs to determine the photons' polarizations. In other words, only a fraction of conversion pairs form an azimuthally modulated ``signal'' which we term $S$, and they sit on top of an unmodulated ``background'' which we call $B$. To quantify the experimental sensitivity to the polarization, we thereby estimate $S/\sqrt{B}$ to be $S/\sqrt{B} = \sqrt{\epsilon_{\rm{cut}}N_{\gamma\text{-}\rm{conv}}} A(\epsilon_{\rm{cut}})$.

This shows the different scaling with the cut efficiency and the analyzing power. This tradeoff is shown in the right panel of Fig.~\ref{fig:convprobandcut}, which displays the figure of merit $\sqrt{\epsilon_{\rm cut}}A$. For double conversion, the estimation scales quadratically with the same variable. We select three benchmark cuts enforcing an energy balance of $\beta_E=0.8$ (enough for both tracks to not be too soft in the detector) and three different angular resolution cuts as shown in Table~\ref{tab:cuts}. In all these selections, we enforce an upper cut of $\theta_\pm<5$ mrad, as events with higher opening angles dilute the analyzing power.

%%%%%%%%%%%%%%%%%%%%%%%%%%%%%%%%%%%%%%%%%%%%%%%%%%%%%%%%%%%%
\begin{table}[h!]
\centering
\begin{tabular}{c | c | c}
\hline\hline
Track resolution & Analyzing power $A$ & Retained fraction $\epsilon_{\rm{cut}}$ \\
\hline
 $0.1$~mrad & $0.248$ & $0.70$ \\
 $0.5$~mrad & $0.193$ & $0.51$ \\
 $1.0$~mrad   & $0.137$ & $0.28$ \\
\hline\hline
\end{tabular}
\caption{Analyzing power $A$ of Si and retained fraction $\epsilon_{\rm{cut}}$ of conversions for three benchmark track angular resolutions. All selections have an upper cut of $\theta_{\pm}<5$ mrad.}
\label{tab:cuts}
\end{table}
%%%%%%%%%%%%%%%%%%%%%%%%%%%%%%%%%%%%%%%%%%%%%%%%%%%%%%%%%%%%

The minimal angular resolution between two tracks is the principal factor in the possibility of extracting polarization information from the pair produced. At these high energies, the effects of multiple scatterings get diluted as the direction of each lepton is weakly affected; the real challenge is reconstructing the separation of the two tracks, which can be a few $\mu$m in the tracker.

%%%%%%%%%%%%%%%%%%%%%%%%%%%%%%%%%%%%%%%%%%%%%%%
\section{Polarization Measurement of a Single Photon}
\label{sec:singlepol}
%%%%%%%%%%%%%%%%%%%%%%%%%%%%%%%%%%%%%%%%%%%%%%%

In this section, we estimate the reach of Belle II for the measurement of the linear polarization of a single converted photon $\gamma_{\rm{conv}}$ to an $e^+ e^-$ pair inside the tracker. For photons originating from $e^+ e^- \to e^+ e^- \gamma_{\rm{conv}}$, we use the selection of highly polarized samples described in Sec.~\ref{sec:eegamma} to estimate the reach for extracting the analyzing power of the detector. For photons from $\gamma \gamma_{\rm{conv}}$, the SM predicts zero linear polarization, and we estimate the smallest polarization that we can exclude. 

We focus on the $e^+e^-$ pair production process. It is also possible that a GeV-energy photon converts into a muon pair, which would offer several advantages, notably an opening angle between the lepton pair that is larger by a factor of $\sim m_\mu/m_e\simeq 200$. However, the probability of this process is highly suppressed compared to the $e^+e^-$ conversion case, by a factor of $(m_e/m_\mu)^2~\sim 2\times10^{-5}$. In either case, these processes provide important backgrounds to searches for displaced vertices from long-lived particles (see, e.g.,~\cite{Jaeckel:2023huy}) and better characterizing them could help to improve the reach for such BSM scenarios.

%%%%%%%%%%%%%%%%%%%%%%%%%%%%%%%%%%%%%%%%%%%%%%%
\subsection{Polarization Measurement from $\gamma \gamma_{\rm{conv}}$}
%%%%%%%%%%%%%%%%%%%%%%%%%%%%%%%%%%%%%%%%%%%%%%%

In the Standard Model (SM), as seen in Eq.~(\ref{eq:Bgammagamma}), the photons in $e^+e^-\to \gamma \gamma$ are individually unpolarized up to the chirality suppression of $m_e^2/s\sim10^{-8}$. Interactions beyond the SM (BSM) could interfere with the SM and could, in principle, induce a nonzero linear polarization. We therefore estimate the sensitivity for a measurement of the degree of linear polarization using $\gamma \gamma$ events at Belle II where one of the two photons converts to an $e^+e^-$ pair (which we term $\gamma \gamma_{\rm{conv}}$~\footnote{The reconstruction of this displaced signature is necessary for the proper background estimation in monophoton searches at Belle~\cite{Essig:2013vha,Graham:2021ggy,Batell:2022dpx,Acanfora:2023gzr,Corona:2024xsg,deLima:2025pzd}. }). 

By using the $\theta_{\rm lab}$-dependent conversion probability, we can estimate the number of single photon conversions we expect as a function of $\theta_{\rm lab}$ by applying a cut on the $e^+e^-$ opening angle of $\theta_{\pm}>0.1$, $0.5$, or $1.0~\rm mrad$. Then, assuming a flat distribution in the azimuthal distribution of the $e^+e^-$ pair, we can estimate the sensitivity to $B_x$ assuming $B_y=0$ using the \textsc{Geant4}-computed analyzing power for each $\theta_{\pm}$ value cut from Table~\ref{tab:cuts}. We show the resulting $2\sigma$ upper limit in each $5^\circ$ $\theta_{\rm lab}$ bin in Fig.~\ref{fig:single_conv} for each choice of opening angle cut. Because the cross-section and the conversion probability peak toward the beamline, the sensitivity is largest at the edges. Since the number of conversions increases as the opening angle cut is loosened, softer requirements on the opening angle lead to stronger limits. A modulation above the SM floor would signal new physics interfering with the two-photon amplitude.

%%%%%%%%%%%%%%%%%%%%%%%%%%%%%%%%%%%%%%%%%%%%%%%
\subsection{Polarization Measurement from $e^+ e^- \gamma_{\rm{conv}}$}
\label{sec:eegammaconv}
%%%%%%%%%%%%%%%%%%%%%%%%%%%%%%%%%%%%%%%%%%%%%%%

%%%%%%%%%%%%%%%%%%%%%%%%%%%%%%%%%%%%%%%%%%%%%%%%%%%%%%%%
\begin{figure}[b!]
    \centering
    \includegraphics[width=1\linewidth]{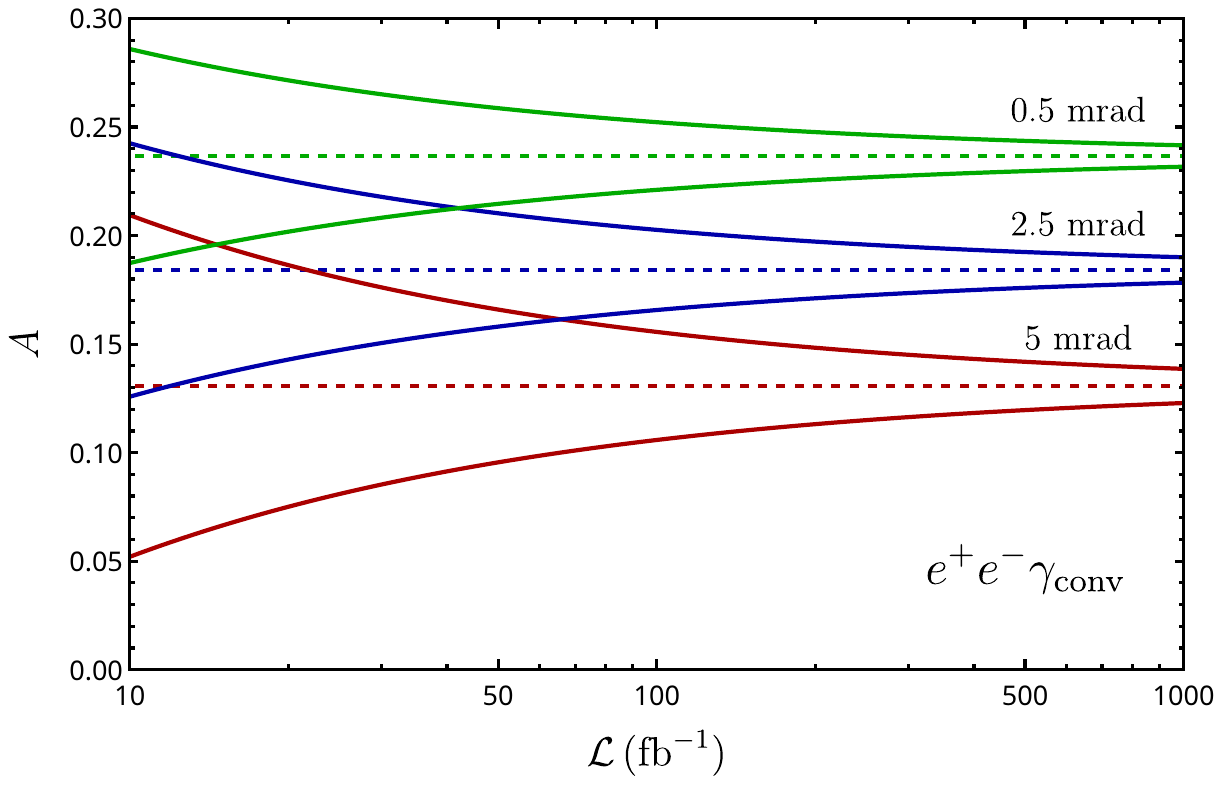}
    \caption{Extraction of the analyzing power using $e^+e^-\gamma_{\rm{conv}}$ events, where the curves show the $\pm2\sigma$ band. This analysis assumes the SM prediction for the Stokes parameters and combines the reach from both the enriched $B_x$ and $B_y$ samples.}
    \label{fig:A_single}
\end{figure}
%%%%%%%%%%%%%%%%%%%%%%%%%%%%%%%%%%%%%%%%%%%%%%%%%%%%%%%%

Because the sample of $e^+e^-\to \gamma \gamma$ events has a negligible amount of linear polarization, it cannot be used to measure the spin analyzing power of the Belle detector. For that, we want to show that it is possible to use the photon conversions of $e^+e^-\to e^+ e^- \gamma_{\rm{conv}}$, which we showed in Sec.~\ref{sec:eegamma} can be selected to contain samples with enriched linear polarization of $B_x$ or $B_y\sim 1$. 

To do this, we estimate the value of the azimuthal modulation of the conversion $e^+e^-$ pair by using the $B_x$ and $B_y$ values of the selections described in Sec.~\ref{sec:eegamma} along with the expected analyzing power from \textsc{Geant4} for different values of the $\theta_{\pm}$ cut. In this case, the lab energy of the photons in the selections is restricted to 1-2 GeV, which makes the opening angle slightly larger than for photons at $\sim 5\GeV$ as in the $\gamma\gamma$ case. We use the same selections as in the $\gamma\gamma_{\rm conv}$ case but, since the opening angle scales as $\theta_{\pm}\propto 1/E_\gamma$, we relax the $\theta_{\pm}$ cut from the $\gamma\gamma$ case to 0.5, 2.5, and $5~\rm mrad$, which keeps the expected analyzing power similar to the values reported in Table~\ref{tab:cuts}.

We use the two samples of events from Sec.~\ref{sec:eegamma} which have different distributions of $B_x$ and $B_y$, along with the selections above to generate azimuthal distributions of $\phi_\pm$. The resulting distributions for the two selections can be seen in the right panel of Fig.~\ref{fig:single_conv} assuming $100~{\rm fb}^{-1}$ of data. The amplitudes of these oscillations are sensitive to $A\times B_{x,y}$. Assuming the values for $B_x$ and $B_y$ we computed in each sample, we can then estimate the sensitivity to extracting a nonzero measurement of the analyzing power $A$ as a function of the amount of data collected. We plot the resulting reach for the analyzing power as a function of luminosity in Fig.~\ref{fig:A_single}. The analyzing power is material dependent, which, in our case, can be either Si or Be. Thus, the analyzing power $A$ obtained by combining all the data and plotted in Fig.~\ref{fig:A_single} is a weighted average of their respective analyzing powers. This reach can be improved with a combination of additional selections of the data to increase the statistics and with further optimization of the angular cut, given the real experimental resolution. This measurement can be made even more robust by using techniques similar to those proposed in~\cite{Cheng:2026ktp} to obtain an independent measurement of the expected polarization of the sample by studying nearly on-shell photons at the same collider. In any case, this analysis shows that photon conversions from $e^+e^-\to e^+e^-\gamma$ events can be used to calibrate the analyzing power without having to fully rely on detector simulations.

%%%%%%%%%%%%%%%%%%%%%%%%%%%%%%%%%%%%%%%%%%%%%%%
\section{Double Conversion and Quantum Entanglement of GeV Photons at Belle II}
\label{sec:double}
%%%%%%%%%%%%%%%%%%%%%%%%%%%%%%%%%%%%%%%%%%%%%%%

When both photons convert, the joint azimuthal distribution of the two pairs measures the correlation between their polarizations. The double-conversion probability is $5\times10^{-4}$, so at $10$ ab$^{-1}$ there are about $10^7$ double conversions in the acceptance before the analyzing-power selection. In this section, we estimate the reach of measurement of the diphoton correlation and several quantum information observables.

%%%%%%%%%%%%%%%%%%%%%%%%%%%%%%%%%%%%%%%%%%%%%%%%%%%
\begin{figure}[t!]
    \centering
    \includegraphics[width=1\linewidth]{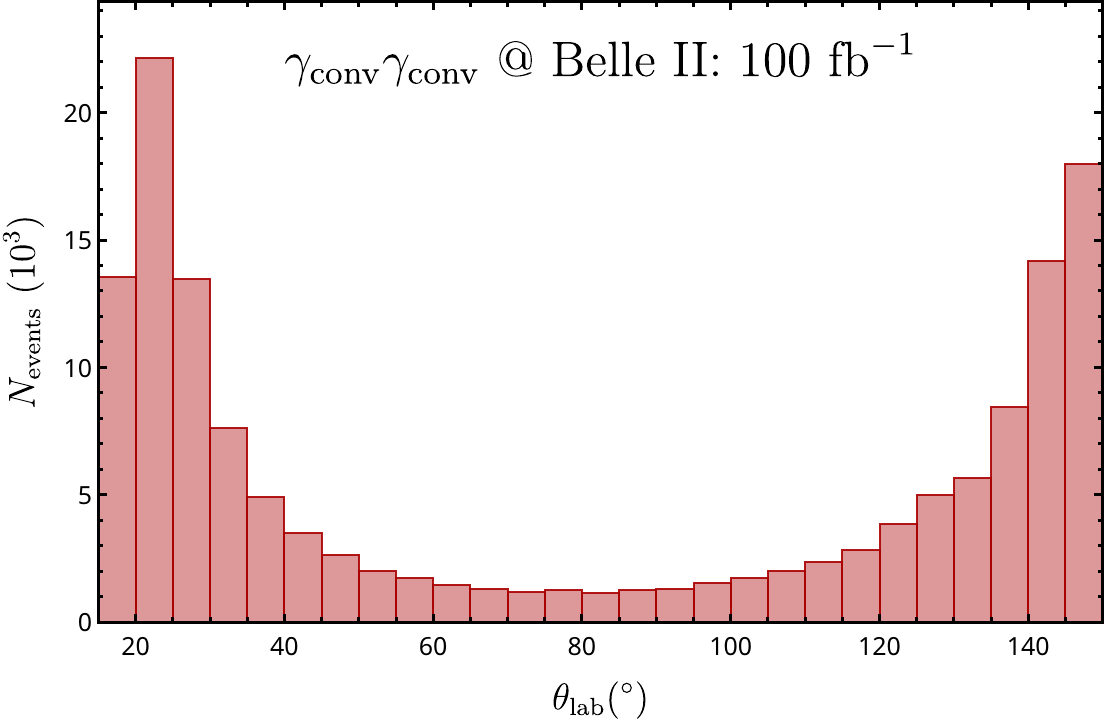} 
    \caption{Expected number of double conversions per $5^\circ$ bin in $\theta_{\rm lab}$ at $100$ fb$^{-1}$. The yield is largest at the edges, where the conversion probability and cross section are highest.}
    \label{fig:ndouble}
\end{figure}
%%%%%%%%%%%%%%%%%%%%%%%%%%%%%%%%%%%%%%%%%%%%%%%%%%%

%%%%%%%%%%%%%%%%%%%%%%%%%%%%%%%%%%%%%%%%%%%%%%%%%%%
\begin{figure*}[t!]
    \centering
     \includegraphics[width=0.45\linewidth]{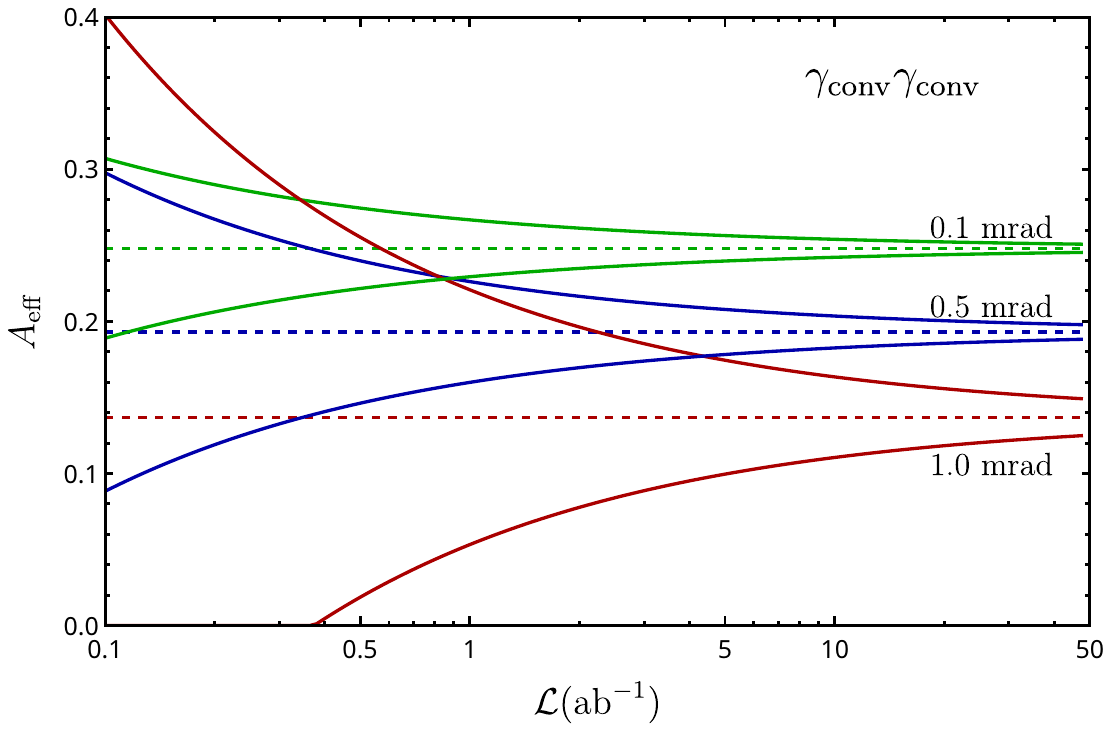}
     \includegraphics[width=0.45\linewidth]{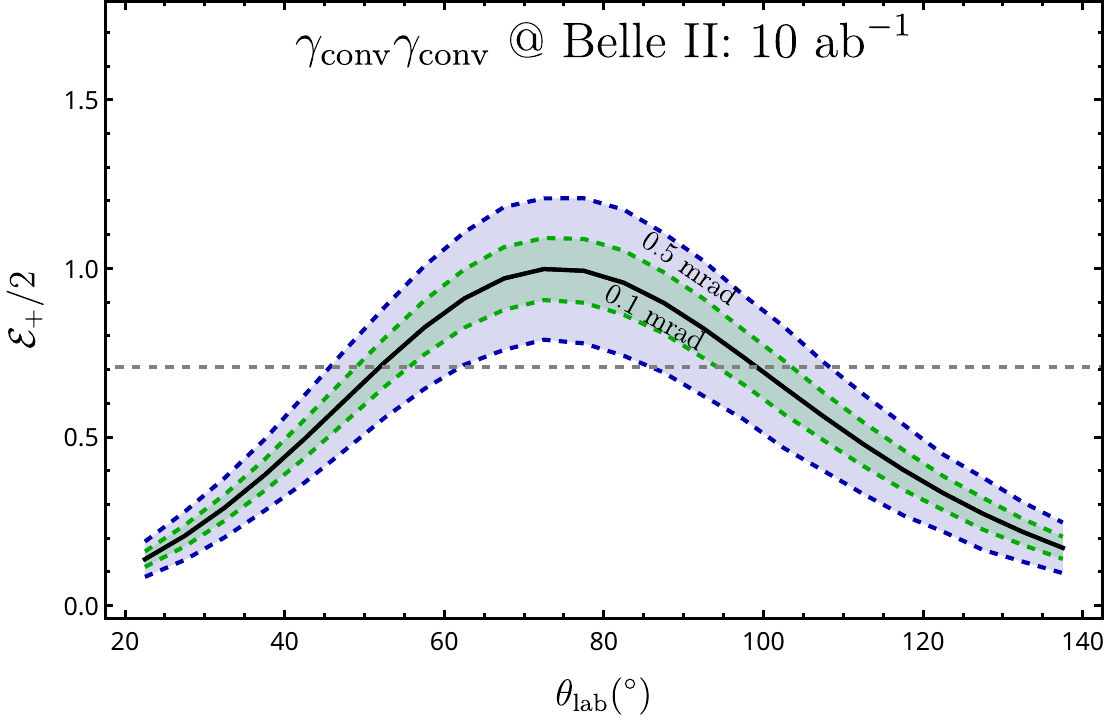}
    \caption{\textbf{Left:} Measured $A_{\rm eff}$ and its $\pm2\sigma$ band as a function of integrated luminosity, assuming the SM correlations. \textbf{Right:} Measurement of the linear polarization correlation $\mathcal{E}_{+}/2=|C_{xx}|=|C_{yy}|$ as a function of $\theta_{\rm lab}$ at $\mathcal{L}=10$ ab$^{-1}$, with the one sigma uncertainty band, assuming the expected analyzing power. Note that quantum mechanics restricts the highest value to be unity, and entanglement is witnessed for values above $1/\sqrt{2}$, shown as a dashed gray line.}
    \label{fig:reach_lum}
\end{figure*}
%%%%%%%%%%%%%%%%%%%%%%%%%%%%%%%%%%%%%%%%%%%%%%%%%%%

%%%%%%%%%%%%%%%%%%%%%%%%%%%%%%%%%%%%%%%%%%%%%%%
\subsection{Photon Correlation Measurement}
\label{sec:reach}
%%%%%%%%%%%%%%%%%%%%%%%%%%%%%%%%%%%%%%%%%%%%%%
The diphoton correlation matrix that appears in the azimuthal distributions of Eq.~\eqref{eq:joint_all} is diagonal in the Stokes basis, with $C_{xx}=C_{yy}$ and $C_{zz}=-1$. The double conversion then predicts a modulation in the two conversion azimuths,
\begin{equation}
\frac{dN}{d(\varphi_1-\varphi_2)}\propto 1+A_1A_2\frac{\mathcal{E}_+}{2}\cos2(\varphi_1-\varphi_2)\, ,
\label{eq:diff_mod}
\end{equation}
with $\mathcal{E}_+/2=C_{xx}=-\sin^2\theta^*/(1+\cos^2\theta^*)$ the linear polarization correlation. The analyzing powers $A_{1,2}$ are evaluated at the respective photon lab energies, between $4.1$ and $6.9\GeV$ through the boost of Eq.~\eqref{eq:boost}, and given the detector material at the conversion point. Because both the cross section and the conversion probability peak at forward angles where the correlation is reduced, the most strongly correlated photon pairs are the rarest. In Fig.~\ref{fig:ndouble}, the yield of double conversion events in the tracker and beam pipe as a function of $\theta_{\rm lab}$ for 100 fb$^{-1}$ is shown. 

%%%%%%%%%%%%%%%%%%%%%%%%%%%%%%%%%%%%%%%%%%%%%%

\begin{table}[b!]
\centering
\begin{tabular}{c| c| c| c| c| c| c| c}
\hline\hline
Conversion [\%]   & \, L$_0$  \, &  \,L$_1$  \, & \, L$_2$  \, & \, L$_3$ \,  & \, L$_4$ \,  & \, L$_5$ \, &  \ L$_6$ \, \\
\hline
 L$_0$ & $2.67$ &        &        &        &        &        &        \\
 L$_1$   & $1.45$ & $0.20$ &        &        &        &        &        \\
 L$_2$   & $1.45$ & $0.39$ & $0.20$ &        &        &        &        \\
 L$_3$   & $6.16$ & $1.67$ & $1.67$ & $3.55$ &        &        &        \\
 L$_4$   & $6.13$ & $1.66$ & $1.66$ & $7.06$ & $3.51$ &        &        \\
 L$_5$   & $6.10$ & $1.65$ & $1.65$ & $7.03$ & $6.99$ & $3.48$ &        \\
 L$_6$   & $6.07$ & $1.65$ & $1.64$ & $6.99$ & $6.96$ & $6.92$ & $3.44$ \\
\hline\hline
\end{tabular}
\caption{Double-conversion layer-pair fractions in the Belle~II vertex detector, single-counted. Values are percentages of all double conversions and sum to $100\%$.}
\label{tab:double_triangle}
\end{table}
%%%%%%%%%%%%%%%%%%%%%%%%%%%%%%%%%%%%%%%%%%%%%%

Two statistical tests can be constructed for this data. Analogously to the procedure of Sec.~\ref{sec:eegammaconv}, we can fix the correlation to the SM value, and then the amplitude returns the effective analyzing power $A_{\rm eff}=\sqrt{A_1 A_2}$, which can be viewed as a test of measuring $A_{\rm eff}\neq 0$. We show the expected precision as a function of luminosity in the left panel of Fig.~\ref{fig:reach_lum} given cuts on the $e^+e^-$ opening angle of $\theta_{\pm}>0.1$, $0.5$, and $1.0~\rm mrad$. As can be seen by comparing to Fig.~\ref{fig:A_single}, the same analyzing powers that enter this quantity can be extracted more easily using single-photon conversions.

Turning this around, we can instead use the measured analyzing powers to extract the correlation between the two photons, characterized by $\mathcal{E}_+/2$ in Eq.~(\ref{eq:diff_mod}). We show the resulting extraction as a function of $\theta_{\rm lab}$ in the right panel of Fig.~\ref{fig:reach_lum}. 

As an aside, we note that the double-conversion process is a background for pairs of displaced vertices that can occur in BSM scenarios where long-lived particles that decay to electrons are produced in pairs.~\footnote{Pair production of long-lived particles can in some sense be more natural than single production, if the long lifetime is associated with a symmetry that is conserved in the production process but weakly broken in decay, which is reminiscent of old puzzles involving strong pair production of long-lived particles carrying strangeness.}   The layer-pair structure of double conversions and their respective materials are given in Table~\ref{tab:double_triangle}.

%%%%%%%%%%%%%%%%%%%%%%%%%%%%%%%%%%%%%%%%%%%%%%%
\subsection{Quantum Information Measurements}
\label{sec:qinfo}
%%%%%%%%%%%%%%%%%%%%%%%%%%%%%%%%%%%%%%%%%%%%%%%
%%%%%%%%%%%%%%%%%%%%%%%%%%%%%%%%%%%%%%%%%%%%%%
\begin{table}[b!]
\centering
\begin{tabular}{ l |c |c |c |c }
\hline
 & Central Value & $\sigma(0.1~\mathrm{mrad})$ & $\sigma(0.5~\mathrm{mrad})$ & $\sigma(1.0~\mathrm{mrad})$ \\
\hline
$\mathcal E_{+\, \rm cen}$      & 1.88  & 0.07  & 0.16  & 0.57  \\
$\mathcal S_{\rm cen}$      & 1.921 & 0.005 & 0.010 & 0.037 \\
\hline
$\mathcal D_{\rm full}$     & 0.251 & 0.013 & 0.031 & 0.110 \\
$\mathcal M_{2,{\rm full}}$ & 0.262 & 0.004 & 0.009 & 0.031 \\
\hline
\end{tabular}
\caption{Central values and statistical uncertainties $\sigma$ for different angular resolutions at $\mathcal L=10~{\rm ab}^{-1}$ for different quantum information observables described in Sec.~\ref{sec:QI}.}
\label{tab:results}
\end{table}
%%%%%%%%%%%%%%%%%%%%%%%%%%%%%%%%%%%%%%%%%%%%%%
Extracting information about quantum correlations between pairs of converted photons produced in $e^+e^-$ collisions requires a good understanding of the analyzing power of photon conversions in different materials in the detector, as well as the conversion probability itself. Since the conversion probability rises toward the acceptance edges, the double conversion weights forward scattering more heavily than the cross section itself, introducing distortions in observables like the correlation matrix $C$ that need to be corrected for when being extracted. Our analysis of  $e^+e^-\to e^+e^-\gamma_{\rm conv}$ shows that many of these difficulties can in principle be mitigated by measurement to reduce systematic uncertainties.

The measurement of the quantum correlation between the photons can be used to reconstruct several quantum information variables described in Sec.~\ref{sec:QI}. We perform two selections for the maximization of the sensitivity of the observables. We consider a central selection, which maximizes quantities that are sensitive to the photons being in pure Bell states, of $58^\circ<\theta_{\rm lab}<91^\circ$. Another selection considers a larger angular region, while avoiding regions too close to the edges where there will be more systematics, of $30^\circ<\theta_{\rm lab}<130^\circ$.

Assuming the data realizes the SM expectation, Table~\ref{tab:results} shows the expected central values and uncertainties of the measurements at $\mathcal L=10~{\rm ab}^{-1}$ for opening angle cuts of $\theta_{\pm}>0.1$, $0.5$, and $1.0~\rm mrad$. The violation of the CHSH inequality of Eq.~(\ref{eq:CHSH_simp}) can be established at $6\sigma$ if a cut of $\theta_{\pm}>0.1$~mrad can be achieved. The steerability, discord, and the magic are measured with relative precisions of $0.26\%$, $5.2\%$, and $1.5\%$ at the best opening angle resolution. 

The method proposed in Ref.~\cite{Cheng:2026ktp} using nearly on-shell virtual photons provides a complementary measurement of these quantities in the same experiment, but with qualitative differences. In our setup, the (real) photons we consider are centimeters apart when they are measured via conversion, making them in principle more sensitive to non-locality provided that timing resolution is sufficient. A loophole-free test of Bell nonlocality is challenging~\cite{Bechtle:2025ugc,Abel:2025skj} and often requires a dedicated experiment~\cite{Clarke:2026vwr}. The main limiting factor of the real photon conversion is the dilution of the analyzing power that comes from the nuclear recoil~\cite{Gros:2016dmp}.

%%%%%%%%%%%%%%%%%%%%%%%%%%%%%%%%%%%%%%%%%%%%%%%
\section{Conclusion}
\label{sec:conc}
%%%%%%%%%%%%%%%%%%%%%%%%%%%%%%%%%%%%%%%%%%%%%%%

We have presented a framework to extract photon polarization and quantum correlations at Belle~II using converted photons in the vertex detector. The photon pair conversion in the detector acts as an imperfect linear polarimeter, and we modeled its response using the five-dimensional polarized Bethe-Heitler simulation in \textsc{Geant4}. We demonstrate that the dominant limitation is the reconstruction of the small opening angle of the $e^+e^-$ pair, which sets the achievable analyzing power and determines the measurement's sensitivity. The proper reconstruction of these pairs will rely on the proper vertex reconstruction, finding the production point inside the beam pipe or tracker material.

Single-conversion events provide a direct way of calibrating the analyzing power of the Belle II detector using polarized photon samples from $e^+e^- \to e^+e^-\gamma$ events. This enables control over detector effects and complements the precision measurement of quantum correlations. Double-conversion events probe the polarization correlation of the diphoton system through joint azimuthal observables. With the expected luminosity of Belle II, these measurements are sensitive to nonclassical correlations and allow for tests of Bell inequalities as well as the extraction of quantum information observables such as discord, magic, and steerability. In particular, we find that these observables can be measured at the percent level or better. Interestingly, the measurements we propose are quantum tests performed at high-energy colliders with quanta that are macroscopically separated when measured. We also note that a Super Tau-Charm Facility~\cite{Ai:2025xop} could probe diphoton correlations at a lower center-of-mass energy where the production cross section and the pair opening angle are larger.

The realistic implementation of this search depends heavily on the vertex reconstruction algorithm. A more complete treatment of detector response, including full reconstruction and backgrounds, would refine the projected reach. Improvements in track reconstruction, particularly for nearly collinear pairs, could enhance the analyzing power and extend the sensitivity. Additionally, exploring additional physical processes, including radiative decays and potential new-physics channels, may provide complementary sources of polarized photons. Specifically, this can offer a new avenue to search for new physics if there are signatures that appear more prominently in the photon polarization and correlations. 

Another avenue to pursue is to consider photon polarization measurements in the design of future detectors, as recently explored in Refs.~\cite{Defurne:2025dkv,Lv:2026qmu} where some proposed designs obtained simulated analyzing power of roughly $0.4$ for a $5~\rm GeV$ photon. This could potentially increase the experimental sensitivity for the observables that we considered, opening up the possibility of Belle nonlocality tests in a collider setting.

It is worth mentioning that this measurement would be the first, to our knowledge, high-precision extraction of the analyzing power of Si and Be for high-energy photons. This measurement can be useful for reducing material systematics in gamma ray polarimeter experiments and testing the high-energy prediction with incredible precision.

%%%%%%%%%%%%%%%%%%%%%%%%%%
\section*{Acknowledgments}
%%%%%%%%%%%%%%%%%%%%%%%%%%
We thank Christopher Hearty for helpful conversations. The work of C.H.L. and D.M. was supported by the Natural Sciences and Engineering Research Council of Canada (NSERC) and by TRIUMF, which receives federal funding via a contribution agreement with the National Research Council (NRC) of Canada. This work was initiated at the Aspen Center for Physics during the workshop "Collider Physics at the LHC and Beyond", which is supported by a grant from the Simons Foundation (1161654, Troyer) and National Science Foundation grant PHY-2210452. 
The work of NM is supported in part by the U.S.~Department of Energy under grant No.~DEFG02-13ER41976/DE-SC0009913, and the DOE QuantISED program through the theory consortium ``Intersections of QIS and Theoretical Particle Physics'' at Fermilab (FNAL 20-17) under contract No.~89243024CSC000002. NM also acknowledges the hospitality of Argonne National Laboratory and the University of Chicago, where part of this research was performed.

\bibliography{ref}

\end{document}